\documentclass[11pt, a4paper, logo, copyright, nonumbering]{fdtn}

\usepackage[numbers, sort]{natbib} 
\usepackage{dblfloatfix}
\usepackage{ulem}
\usepackage{caption}
\usepackage{dramatist}
\usepackage{xspace}

\usepackage{multirow}
\usepackage{xltabular}
\usepackage{longtable}
\usepackage{hyperref}
\usepackage{amsfonts}
\usepackage{amsmath}
\usepackage{amssymb}
\usepackage{lineno}
\usepackage{adjustbox}

\usepackage[bottom]{footmisc}

\usepackage{subfigure}
\usepackage{setspace}

\usepackage{graphicx}
\usepackage{array, booktabs, tabularx}
\hypersetup{hidelinks=true, colorlinks=true, citecolor=blue, linkcolor=black, urlcolor=black}
\usepackage[textsize=scriptsize]{todonotes}
\usepackage{algorithm}
\usepackage{algpseudocode}
\usepackage{listings}
\usepackage{xcolor}
\usepackage{needspace}
\Needspace{8\baselineskip}

\usepackage{makecell}
\usepackage{inconsolata}

\usepackage{fvextra}
\usepackage{pgfplots}
\pgfplotsset{compat=1.18}
\usepackage{pgf-pie}
\usepackage{fancyvrb}
\usepackage{mdframed}

\usepackage{tcolorbox}
\usepackage{arydshln}
\usepackage{threeparttable}
\usepackage{subcaption}
\usepackage{wrapfig}

\usepackage{dsfont}
\usepackage{array}
\usepackage{tabularx}
\usepackage{xcolor}

\usepackage{lipsum}
\usepackage{multicol}

\usepackage[utf8]{inputenc}
\usepackage[T1]{fontenc}  
\usepackage{hyperref}       
\usepackage{url}            
\usepackage{booktabs}       
\usepackage{amsfonts}       
\usepackage{nicefrac}       
\usepackage{microtype}      
\usepackage{xcolor}         

\usepackage{environ}      
\usepackage{tcolorbox}
\tcbuselibrary{listings,breakable}

\makeatletter
\def\@BTrule[#1]{
  \ifx\longtable\undefined
    \let\@BTswitch\@BTnormal
  \else\ifx\hline\LT@hline
    \nobreak
    \let\@BTswitch\@BLTrule
  \else
     \let\@BTswitch\@BTnormal
  \fi\fi
  \global\@thisrulewidth=#1\relax
  \ifnum\@thisruleclass=\tw@\vskip\@aboverulesep\else
  \ifnum\@lastruleclass=\z@\vskip\@aboverulesep\else
  \ifnum\@lastruleclass=\@ne\vskip\doublerulesep\fi\fi\fi
  \@BTswitch}
\makeatother

\addto\extrasenglish{

}

\reportnumber{001} 

\newcommand{\PretrainedModelName}[0]{Foundation-Sec-8B} 
\newcommand{\IFTModelName}[0]{Foundation-Sec-8B-Instruct} 

\newcommand{\llamathree}{Llama 3.1}

\newcommand{\gptthreefive}{GPT-3.5}

\newcommand{\gptfourOmini}{GPT-4o-mini}

\newcommand{\primus}{Primus}

\newcommand{\qwen}{Qwen-2.5}
\newcommand{\tulu}{Tülu}
\newcommand{\tuluthree}{\tulu{} 3}
\newcommand{\tulutwo}{\tulu{} 2}
\newcommand{\gemma}{Gemma-3}
\newcommand{\whiterabbitneo}{DeepHat-v1} 
\newcommand{\lily}{LilyCybersecurity}

\newcommand{\ctibench}{CTIBench}
\newcommand{\seceval}{SecEval}
\newcommand{\secbench}{SecBench}
\newcommand{\cybermetric}{CyberMetric}

\newcommand{\gsmk}{GSM8K}
\newcommand{\ifeval}{IFEval}
\newcommand{\matheval}{MATH}
\newcommand{\mmlu}{MMLU}
\newcommand{\mmlusec}{MMLU-Computer Security}
\newcommand{\bbh}{BigBenchHard}

\newcommand{\humaneval}{HumanEval}
\newcommand{\alpacaleval}{AlpacaEval 2}
\newcommand{\bigbench}{BIG-bench}

\newcommand{\wizardevol}{WizardLM Evol Instruct}
\newcommand{\nemotron}{Nemotron SFT}
\newcommand{\nemotronchat}{\nemotron{} Chat}
\newcommand{\nemotronsafety}{\nemotron{} Safety}
\newcommand{\alpaca}{Alpaca}
\newcommand{\openplatypus}{Open Platypus}
\newcommand{\openorca}{OpenOrca}

\newcommand{\olmes}{OLMES}

\renewcommand{\phi}{\varphi}

\renewcommand{\leq}{\leqslant}
\renewcommand{\geq}{\geqslant}

\renewcommand{\epsilon}{\varepsilon}
\renewcommand{\imath}{\mathrm{i}}

\newlength{\restsubwidth}
\newlength{\restsubheight}
\newlength{\restsubmoreheight}
\newcommand{\rest}[2]{%
        \settowidth{\restsubwidth}{\ensuremath{#2}}
        \settoheight{\restsubheight}{\ensuremath{{}_{#2}}}
        \ensuremath{{#1\hskip 0.5pt}_{\vrule\kern2pt\parbox[b][%
        4pt][b]{\the\restsubwidth}{%
                        \ensuremath{{}_{#2}}}}}
        }

\definecolor{customlink}{HTML}{A9336A}
\renewcommand{\today}{}

\title{Llama-3.1-FoundationAI-SecurityLLM-8B-Instruct Technical Report}

\author{
    Sajana~Weerawardhena$^{1}$,
    Paul~Kassianik$^{1}$,
    Blaine~Nelson$^{1}$, 
    Baturay~Saglam$^{1, 2, *}$, 
    Anu~Vellore$^{1}$, 
    Aman~Priyanshu$^{1}$, 
    Supriti~Vijay$^{1, 3, *}$, 
    Massimo~Aufiero$^{1}$,
    Arthur~Goldblatt$^{1}$,
    Fraser~Burch$^{1}$,
    Ed~Li$^{1, 2, *}$,
    Jianliang~He$^{1, 2, *}$,
    Dhruv~Kedia$^{1}$,
    Kojin~Oshiba$^{1}$,
    Zhouran~Yang$^{1,2, *}$,
    Yaron~Singer$^{1}$,
    Amin~Karbasi$^{1}$\\
    \vspace{0.5em}
    {\normalsize $^{1}$Foundation AI--Cisco Systems Inc.} \\
    {\normalsize $^{2}$Yale University} \\
    {\normalsize $^{3}$Carnegie Mellon University} \\
    \vspace{1.0em}
    {\small $^{*}$Work done while at Foundation AI} \\
}

\begin{abstract}

Large language models (LLMs) have shown remarkable success  across many domains, yet their integration into cybersecurity applications remains limited due to 
a lack of general-purpose cybersecurity data, representational complexity, and safety and regulatory concerns.
To address this gap, we previously introduced \PretrainedModelName{}, a cybersecurity-focused LLM suitable for fine-tuning on downstream tasks. 
That model, however, was not designed for chat-style interactions or instruction-following. 
In this report, we release \IFTModelName{}: a model specifically trained for general-purpose cybersecurity dialogue. 
Built on \PretrainedModelName{}, it combines domain-specific knowledge with instruction-following, conversational capabilities, and alignment with human preferences to produce high-quality, relevant responses. 
Comprehensive evaluations show that \IFTModelName{} outperforms \llamathree{}-8B-Instruct on a range of cybersecurity tasks while matching its instruction-following performance. 
It is also competitive with \gptfourOmini{} on cyber threat intelligence and instruction-following tasks.
We envision \IFTModelName{} becoming an indispensable assistant in the daily workflows of cybersecurity professionals.
We release the model publicly at \href{https://huggingface.co/fdtn-ai/Foundation-Sec-8B-Instruct}{\textcolor{customlink}{\texttt{https://huggingface.co/fdtn-ai/Foundation-Sec-8B-Instruct}}}.

\end{abstract}

\begin{document}

\maketitle

\section{Introduction}

Frontier large language models (LLMs) are driving innovation across a swath of domains. 
Tools like ChatGPT \citep{chatgpt} have seen widespread use, assisting users in tasks ranging from creative writing to technical problem-solving. 
Specialized models have also shown promise in fields such as medicine \cite{singhal2025toward, meditron}, law \cite{colombo2024saullm, colombo2024saullm54b}, and code generation \cite{codellama}. 
Nevertheless, integration into mainstream cybersecurity workflows remains limited. 
Practitioners still face a range of practical challenges while commercial models often impose strict safety guardrails that constrain their practical utility for security applications~\citep{cybersec_llms_survey}.

Lack of data, alignment gaps, and field breadth are key challenges to practical deployment of cybersecurity-specific LLMs.
The scarcity of clean, high-quality, public datasets hinders the development of models with robust cybersecurity capabilities \citep{mbona2024datasets}.
Interpretability research on LLM representations is still evolving, leaving models vulnerable to hallucinations \citep{Tonmoy2024-kz} and distribution shifts \citep{Gupta2025-yg}.
The broad scope of cybersecurity topics and tasks---from phishing and malware detection to cryptographic operations---further complicates efforts to build a single, general-purpose model. 
As a result, the ecosystem remains fragmented and dominated by narrow, task-specific tools \citep{zhang2025llms, xu2024large, motlagh2024large}.

To address these challenges, we previously introduced \PretrainedModelName{}, a cybersecurity-specialized LLM based on \llamathree{}-8B \citep{kassianik2025llama}. 
Through continual pre-training on a large cybersecurity corpus, the model is infused with deep domain knowledge and serves as a foundation for general cybersecurity tasks. 
However, as a \textit{base model}, it is not trained to interact with users, follow instructions in natural language, or align responses to human preferences.

In this work, we present \IFTModelName{}, a cybersecurity chat model built on \PretrainedModelName{} with instruction-following and conversational capabilities. Base models alone typically require task-specific fine-tuning \citep{howard2018universal} or carefully designed in-context examples \citep{brown2020language} before they can provide meaningful interactions. \IFTModelName{} makes the knowledge embedded in \PretrainedModelName{} readily accessible—users can query the model directly and engage in interactive dialogue without any prerequisite setup. By combining domain expertise with instruction-tuning and preference alignment, \IFTModelName{} enables a broad range of interactions and real-world workflows.

\begin{figure}[tpb]
\centering
\includegraphics[width=0.8\textwidth]{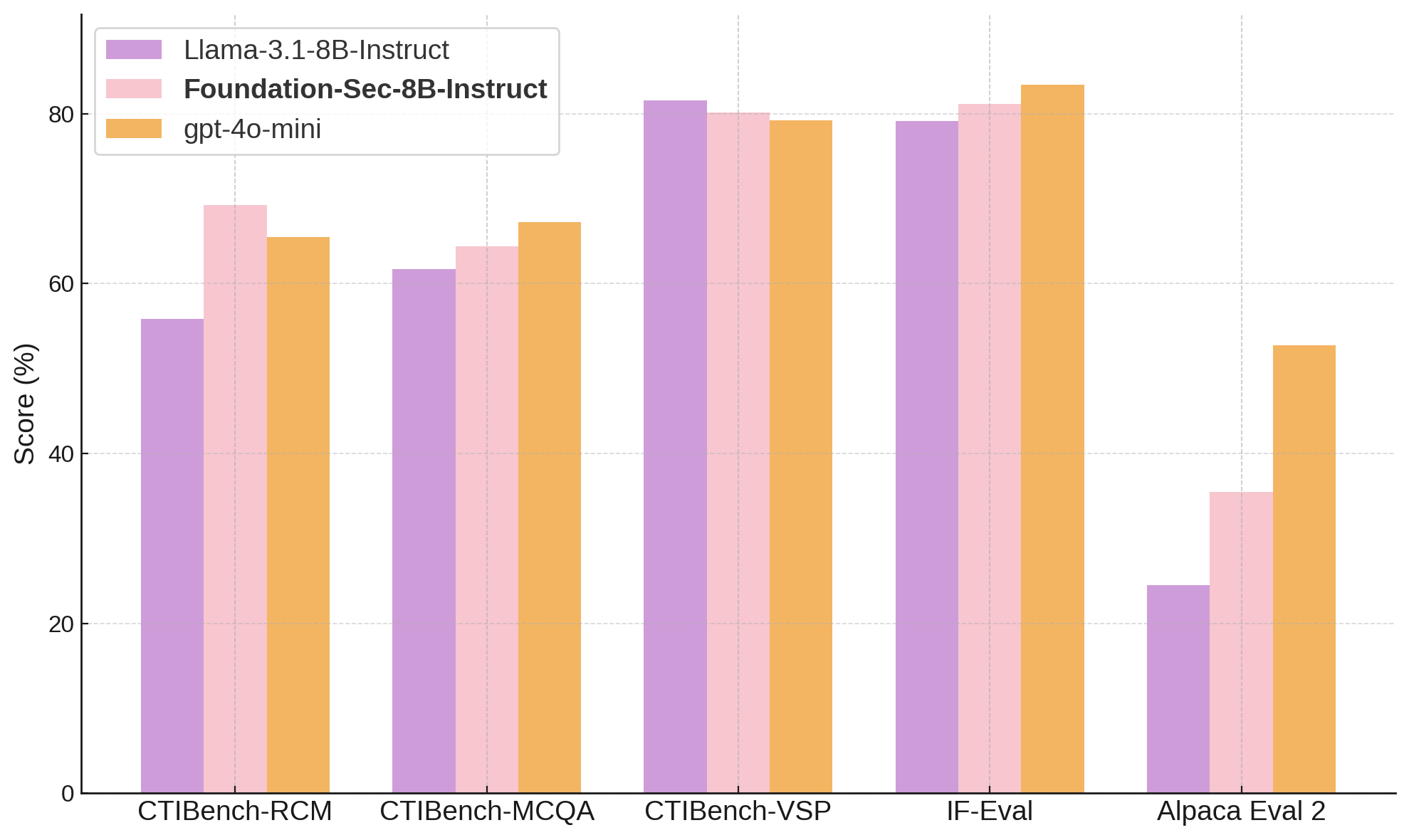}
\caption{Overview of core results on the selected benchmarks. \IFTModelName{} matches or outperforms \llamathree{}-8B-Instruct and \gptfourOmini{} in cyber threat intelligence tasks, and surpasses \llamathree{}-8B-Instruct on instruction-following tasks.}
\label{fig:summary-benchmarks}
\end{figure}

We provide comprehensive evaluations across diverse benchmarks to assess our model in both cybersecurity knowledge and instruction-following ability. 
Results show that \IFTModelName{} outperforms \llamathree{}-8B-Instruct on cyber threat intelligence (CTI) benchmarks and remains competitive with \gptfourOmini{} on both cybersecurity and general instruction-following tasks (see Figure \ref{fig:summary-benchmarks}). 
We also include a safety analysis of the model’s vulnerability to malicious prompts. 
Our goal is for \IFTModelName{} to become a trusted daily assistant for cybersecurity professionals worldwide.

\section{Related Work}

We examine prior work on instruction-tuned, cybersecurity-specialized models and relevant post-training approaches. 
While much of the literature has focused on secure code generation or vulnerability detection \cite{kassianik2025llama}, our goal is to align a general-purpose LLM with core cybersecurity tasks. 
We aim to enable broader applicability of the model across diverse use cases through instruction-following training.
For a comprehensive overview of cybersecurity-focused language models, we refer readers to existing surveys \citep{cybersec_llms_survey}.

\subsection{Instruction-Tuned Cybersecurity LLMs}

To our knowledge, only a few instruction-following models have been developed for general cybersecurity tasks. We include them as baselines in our evaluations.

\paragraph{\whiterabbitneo{} \citep{whiterabbitneo}} A family of instruction-tuned cybersecurity models based on \llamathree{} and \qwen{}.
DeepHat is intentionally uncensored and optimized for offensive security applications, including red teaming and exploit development.

\paragraph{\primus{} \citep{yu2025primus}} An instruction-tuned cybersecurity LLM built on \llamathree-8B, trained on a curated 2 billion-token corpus sourced from MITRE ATT\&CK, Wikipedia, vendor reports, threat intelligence feeds, and manually collected CTI data.

\paragraph{\lily{} \citep{lily}} A cybersecurity assistant based on Mistral-7B \citep{jiang2023mistral7b}, fine-tuned with supervision on 22,000 hand-crafted conversations related to cybersecurity and hacking.

\subsection{Post-Training}

Modern post-training recipes improve model abilities in reasoning, math, coding, and tool use, while also adapting next-token prediction to better follow user instructions \citep{ouyang2022training, instruct_gpt}. Recent approaches have replaced complex multi-stage reinforcement learning (RL) algorithms, such as Proximal Policy Optimization (PPO) \citep{ppo}, with simpler methods like Direct Preference Optimization (DPO) \citep{rafailov2023direct}.

As of this writing, and to our knowledge, there are no high-quality cybersecurity datasets or benchmarks for evaluating alignment with human preferences. In contrast, fields like medicine and law have emphasized the need to assess model outputs against domain-specific human judgments \citep{guha2023legalbench, singhal2025toward}.

Most existing cybersecurity benchmarks, however, focus on knowledge recall rather than alignment with human preferences. 
They tend to emphasize multiple-choice questions \citep{li2023seceval, tihanyi2024cybermetric, liu2023secqa} or narrowly scoped tasks like classification (e.g., vulnerability detection), named entity recognition (NER), and summarization \citep{deka2024attacker, zhang2024cybench}. 
Models evaluated only on such benchmarks leave open the question of whether they are suitable for broader, real-world use \citep{yu2025primus}.
In contrast, our model not only performs well on cybersecurity knowledge recall tasks but also shows general alignment with human preferences, making it better suited for practical applications.

\section{Benchmarks}

\begin{table}[htpb]
  \centering
  \resizebox{0.45\textwidth}{!}{
    \begin{threeparttable}
    \begin{tabular}{lll}
      \toprule
      \textbf{Benchmark}
        & \textbf{Domain}  \\
      \midrule
      \ctibench{}-MCQA
        & cybersecurity \\
      \ctibench{}-RCM
        & cybersecurity  \\
      \ctibench{}-VSP
        & cybersecurity \\
      \cybermetric{}-500
        & cybersecurity  \\
      \secbench{}
        & cybersecurity \\
      \seceval{}
        & cybersecurity \\
      \midrule
      \mmlu 
        & knowledge \\
      \ifeval 
        & instruction following \\
      \gsmk{}
        & math  \\ 
      \matheval
        & math  \\ 
      \humaneval
        & code  \\ 
      \bbh 
        & reasoning \\
       \alpacaleval
        &  human preference \\
      \bottomrule
    \end{tabular}
    \end{threeparttable}
  }
  \caption{Benchmarks used for our model evaluations and comparisons.}
  \label{tab:benchmark-types}
\end{table}

We evaluate \IFTModelName{} on both cybersecurity-specific and general post-training benchmarks. 
We evaluate the model on both cybersecurity-specific and general post-training benchmarks, chosen to assess domain knowledge, instruction-following, practical utility, and reasoning skills. Table~\ref{tab:benchmark-types} summarizes all benchmarks used in our evaluation.

\subsection{Security Benchmarks}
\label{sec:sec_benchmarks}

Our principal assessment focuses on the model’s cybersecurity knowledge. 
We use a broad set of benchmarks that cover security governance and compliance, technical and infrastructure protection, threat detection and incident response, and emerging risks. 
Most benchmarks are formatted as multiple-choice question-answering (MCQA) tasks (i.e., \ctibench{}-MCQ, \cybermetric{}, \secbench{}, \seceval{}, \mmlusec{}), while others (i.e., \ctibench{}-RCM and \ctibench{}-VSP) use custom short-answer formats. 
A full list and detailed descriptions of all security benchmarks are provided in Appendix~\ref{app:sec_benchmarks}.

These benchmarks also evaluate general instruction-following capabilities within a cybersecurity context.
Each security benchmark can be framed as a cybersecurity instruction-following task with verifiable answers.
We use performance on these tasks as a proxy for the model’s instruction-following ability in the security domain.
Sample prompts and answers are provided in Appendix~\ref{app:sec_benchmarks}.

\subsection{Post-Training Evaluations} 
\label{sec:post_training_evals}

We evaluate our model on several general-purpose benchmarks, including Massive Multitask Language Understanding (\mmlu{}) \citep{mmlu}, \ifeval{} \citep{zhou2023instruction}, \gsmk{} \citep{cobbe2021gsm8k}, \matheval{} \citep{hendrycks2021measuring}, \humaneval{} \citep{chen2021evaluating}, AlpacaEval 2 \citep{dubois2024length}, and \bbh{} \citep{suzgun2022challenging}. 
These benchmarks assess instruction-following, coding, math, and reasoning capabilities. 
Detailed descriptions are provided in Appendix~\ref{app:gen_benchmarks}. 
We use the Open Language Model Evaluation System (\olmes{}) \citep{gu2024olmes} for all general benchmarks except \mmlu{}, for which we use our own evaluation framework.

\subsection{Cybersecurity Topics in Evaluations}
To better understand the topic distribution of the cybersecurity benchmarks, we analyzed sub-topic coverage using our own taxonomy; see Appendix~\ref{app:taxonomy} for details.

The analysis in Figure~\ref{fig:benchmark-topic-dist} shows a skewed distribution: areas like application security are heavily overrepresented, while other areas, such as security operations and cloud security, are underrepresented.

\begin{figure}[thpb]
\centering
\includegraphics[width=0.95\textwidth]{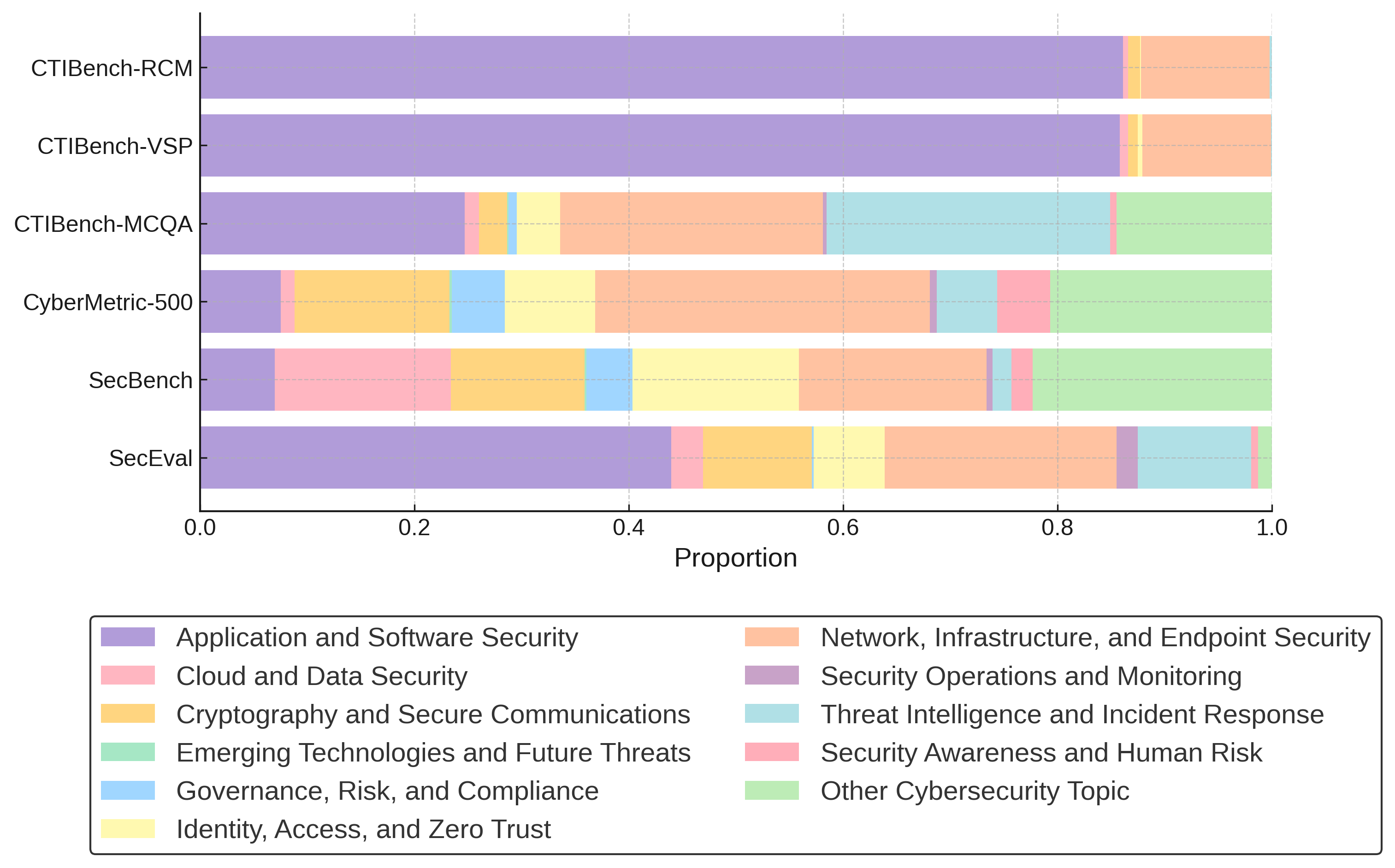}
\caption{Distribution of topics in common cybersecurity evaluation benchmarks. Application security is heavily overrepresented, particularly in \ctibench{} tasks, as well as network security and cloud security.}
\label{fig:benchmark-topic-dist}
\end{figure}

\section{Post-Training Dataset Analysis} \label{sec:post_analysis}

\subsection{Datasets} \label{sec:post_training_dataset_list}

In this section we study distribution of cybersecurity data in common open-sourced post-training datasets. Specifically we study: \tuluthree{} \citep{tulu3}, \tulutwo{} \citep{tulu2}, \alpaca{} \citep{alpaca}, \openplatypus{} \citep{lee2023platypus}, \wizardevol{} (70k) \citep{xu2023wizardlm}, \openorca{} \citep{OpenOrca}, Llama-Nemotron Post-Training Dataset (non-reasoning, chat, and safety components only) \citep{nemotron}, \primus-Instruct and \primus-Reasoning \citep{yu2025primus}.

\subsection{Analysis Of Cybersecurity Data In Post-Training Datasets} \label{sec:cyberdata}

As part of the data curation process, we studied the presence and makeup of cybersecurity data in the datasets listed in Section~\ref{sec:post_training_dataset_list} (excluding \primus-Instruct and \primus-Reasoning as they are cybersecurity-oriented datasets).

We employ a two-stage classification system to identify and categorize cybersecurity-focused prompts according to our in-house cybersecurity taxonomy described in Appendix~\ref{app:taxonomy}. 
The first stage consists of keyword and regex-based pattern matching rules derived from this taxonomy. 
For additional details on the keyword-based classification stage, see Appendix~\ref{app:post_training_data_analysis}. 
The second stage applies a fine-tuned ModernBERT \citep{modernbert} model to perform multi-class classification in alignment with our taxonomy.

\begin{table}[h]
\centering
\begin{tabular}{lcccc}
\toprule
\textbf{Source} & \textbf{Estimated Count} & \textbf{Estimated Percentage} \\ 
\midrule
\openorca{} & 8555 & 0.20\%  \\
\tuluthree{} SFT Mixture & 13083 & 1.39\%  \\
\tulutwo{} SFT Mixture & 4309 & 1.32\%  \\
\wizardevol{} & 1198 & 1.71\%  \\
\nemotronchat{} & 711 & 1.79\% \\
\nemotronsafety{} & 170  & 0.68\%  \\
\alpaca{} & 168 &  0.32\%  \\
\openplatypus{} & 100  & 0.32\%   \\
\bottomrule
\end{tabular}
\caption{Estimated count and percentage of cybersecurity-focused prompts in open-sourced post-training datasets.}
\label{tab:keyword_based_filter_lower_upper_bounds}
\end{table}

Table~\ref{tab:keyword_based_filter_lower_upper_bounds} summarizes our findings. 
Several datasets exhibit minimal security content with estimated ranges below 1\%, but the \wizardevol{}, \nemotronchat{}, and \tulu{} datasets likely have non-trivial proportions of cybersecurity data.
More comprehensive statistics are provided in Appendix~\ref{app:post_training_data_analysis}.

\begin{figure}[h!]
\centering 
\includegraphics[width=\textwidth]{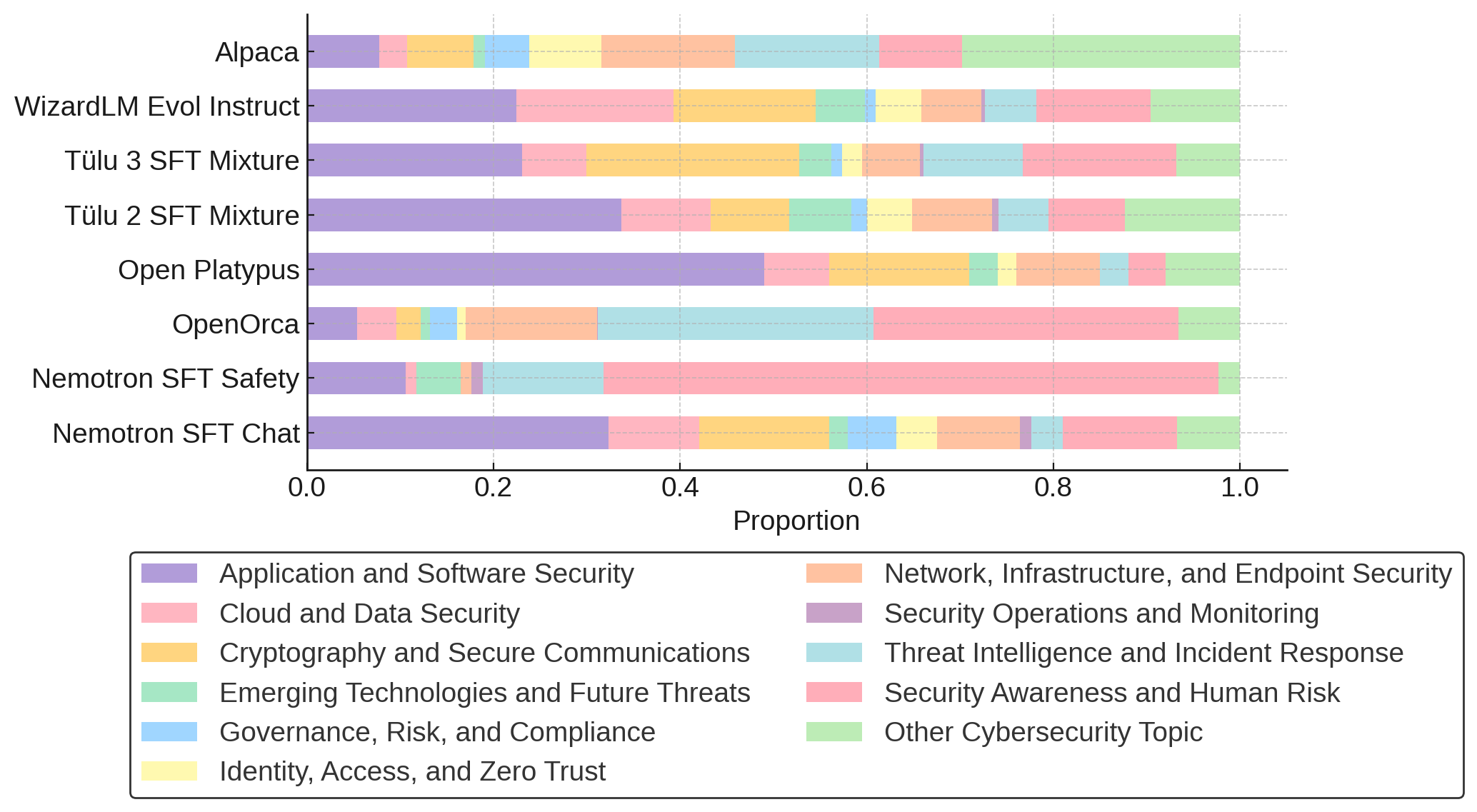}
\caption{Distribution of cybersecurity topics within post-training datasets.}
\label{fig:post-training-dist}
\end{figure} 

Figure~\ref{fig:post-training-dist} shows the breakdown of each dataset by cybersecurity category as determined by our classification model. 
These post-training datasets demonstrate much greater coverage and balanced distributions of categories than the evaluation datasets in Figure~\ref{fig:benchmark-topic-dist}.

This analysis may be of interest to the broader cybersecurity community, offering \textit{insight into the composition of cybersecurity data in open-source post-training datasets}.

\subsection{Understanding Contamination on Post-Training Datasets}

\begin{table*}
\centering
\setlength{\tabcolsep}{4.5pt} 
\renewcommand{\arraystretch}{1.6}
\definecolor{custommustard}{HTML}{ffecec}
\definecolor{lightred}{HTML}{ffcccc}
\definecolor{darkred}{HTML}{ff9797}
\resizebox{\textwidth}{!}{
\begin{tabular}{l *{12}{c}} 
\toprule
  & \multicolumn{4}{c}{\textbf{\tuluthree{} SFT}} & \multicolumn{4}{c}{\textbf{Primus-Reasoning}} & \multicolumn{4}{c}{\textbf{Alpaca}} \\
\cmidrule(lr){2-5} \cmidrule(lr){6-9} \cmidrule(lr){10-13}
Benchmark & EM & NG & ES & \mbox{LLM-ES} & EM & NG & ES & \mbox{LLM-ES} & EM & NG & ES & \mbox{LLM-ES} \\
\midrule
\ctibench-MCQA & 0.00 & 0.00 & \cellcolor{lightred}4.12 & \cellcolor{custommustard}0.1 & \cellcolor{lightred}33.40 & \cellcolor{lightred}32.04 & \cellcolor{lightred}13.48 & \cellcolor{lightred}10.2 & 0.00 & 0.00 & \cellcolor{lightred}3.44 & \cellcolor{custommustard}0.1 \\
\ctibench-RCM & 0.00 & 0.00 & \cellcolor{custommustard}1.70 & 0.00 & \cellcolor{darkred}83.10 & \cellcolor{darkred}78.80 & \cellcolor{darkred}91.50 & \cellcolor{darkred}74.9 & 0.00 & 0.00 & 0.00 & 0.00 \\
\ctibench-VSP & 0.00 & 0.00 & \cellcolor{lightred}3.70 & 0.00 & \cellcolor{darkred}83.10 & \cellcolor{darkred}78.20 & \cellcolor{darkred}78.40 & \cellcolor{darkred}72.6 & 0.00 & 0.00 & \cellcolor{custommustard}0.10 & 0.00 \\
\cybermetric-500 & 0.00 & 0.00 & \cellcolor{lightred}28.00 & \cellcolor{custommustard}0.6 & 0.00 & 0.00 & 0.00 & 0.00 & 0.00 & 0.00 & \cellcolor{lightred}33.20 & \cellcolor{lightred}2.2 \\
\bottomrule
\end{tabular}
}
\caption{Contamination levels across selected post-training datasets relative to benchmark sizes (percentages, rounded). 
EM = Exact Match, NG = N-Gram Overlap, ES = Embedding Similarity, and LLM-ES = LLM-verified Embedding Similarity. 
\colorbox{custommustard}{Light red} cells indicate values $> 0\%$ but $\leq 2\%$; \colorbox{lightred}{red} cells indicate values $> 2\%$ but $\leq 50\%$; \colorbox{darkred}{dark red} cells indicate values $> 50\%$.}
\label{tab:contamination}
\end{table*}

Large-scale post-training datasets are often aggregated from diverse and minimally filtered sources, making them prone to unintentional inclusion of benchmark content \cite{deng-etal-2024-investigating, xu2024benchmarkdatacontaminationlarge}. Such contamination can artificially inflate reported performance, distort comparative evaluations, and undermine our ability to assess a model’s true generalization capabilities \cite{fu-etal-2025-data}. In this analysis, we aim to systematically characterize the extent and nature of contamination across commonly used post-training datasets. Understanding where and how benchmark leakage occurs enables more informed decisions about dataset selection for model training and guides the design of decontamination pipelines that preserve benchmark integrity.

Building on prior work \cite{tulu3, yang2023rethinkingbenchmarkcontaminationlanguage, zhu-etal-2024-inference, deng-etal-2024-investigating, xu2024benchmarkdatacontaminationlarge}, we implement a layered framework comprising three components: an n-gram–based search to detect verbatim matches, an embedding-based cosine similarity filter to identify semantically similar phrases and an LLM-as-a-Judge quality filter to improve the detection of paraphrases. These components provide a robust mechanism for detecting contamination, though results remain sensitive to hyperparameter choices.

We apply this framework to several widely used post-training datasets and summarize their overlap with security-focused benchmarks in Table~\ref{tab:contamination}.
As was reported in its dataset report~\cite{yu2025primus}, the \primus-Reasoning dataset was constructed using cybersecurity reasoning tasks from \ctibench{}.
Our contamination detection correctly identifies this contamination with all three components, particularly in \ctibench{}-RCM (79-91\% detected) and \ctibench{}-VSP (72-83\% detected).
This validates that our approach correctly detects known contamination and highlights the risk posed by training on datasets without conducting contamination audits, as we evaluate on both of these benchmarks.
The contamination of \ctibench{} datasets is low (<5\%) in other datasets, while modest levels of contamination from \cybermetric{}-500 are found in \tuluthree{} and \alpaca{}. 

Notably embedding similarity dominates contamination detection in these datasets. However, embedding similarity between corpora of security data tends to be high by default, which
can inflate contamination estimates. We attribute this to several factors: embedding similarity reflects semantically learned representations, and the relatively limited presence of cybersecurity data in the pre-training of these embedding models may reduce their ability to capture fine-grained distinctions in this domain \cite{9207140, bayer2022cysecbertdomainadaptedlanguagemodel};  moreover, similar to known issues with embedding metrics like BERTScore, these embeddings may conflate shared domain-specific vocabulary with true content overlap \cite{nguyen-etal-2021-cross-domain, sun2022bertscoreunfairsocialbias}. To reduce false positives and improve robustness, we tuned to a higher similarity threshold (e.g., cosine 0.75–0.8) and validated a subset of matches using an LLM-as-a-judge approach. A detailed breakdown of our choices and  representative examples at different similarity levels can be found in Appendix~\ref{app:detailed_decont}.

\section{Training }

\subsection{Cybersecurity Data Mix}

Prior work has suggested that adding new knowledge during fine-tuning increases the likelihood of hallucinations  \citep{gekhman2024does, mielke2022reducing}. We adopt the stance that post-training should not attempt to impart novel cybersecurity knowledge.  Instead, we rely primarily on pre-training of our base model to provide this knowledge.  We further observed that including large amounts of cybersecurity data in our post-training datasets tends to dilute the effectiveness of our knowledge-based cybersecurity evaluations, making it difficult to assess generalization and compare between model candidates. 

\subsection{Supervised Fine-tuning and Direct Preference Optimization}

Supervised fine-tuning is utilized to hone core instruction-following skills in the model. 
In contrast, RL techniques are primarily used to improve on instruction-following and alignment with human preferences. 
Starting from \PretrainedModelName{}, we apply a combination of SFT and RL techniques to develop our final instruction-tuned model.

Similar to other post-training procedures, we also rely on high quality synthetic data \citep{abdin2024phi, grattafiori2024llama, yang2025qwen3}.  
We built a synthetic data generation pipeline with a focus on refinement through rejection sampling, difficulty grading, and automated verification checks. 
Our final model also went through several rounds of human preference testing that revealed insights which were then used to bolster our preference tuning data.

Consistent with prior work, we note that data diversity during post-training drives model generalization \citep{zhou2023lima, wang2022super, chung2024scaling}. 
In a similar vein, we observe that data diversity is also a key driver of knowledge retention. 
Prior work showed that while minor degradation is expected during post-training, the severity can be mitigated by increasing the diversity and balance of the post-training dataset \citep{grattafiori2024llama, olmo2}.
As described in Section~\ref{sec:forgetting}, \IFTModelName{} only shows  mild degradation in knowledge tasks compared to \PretrainedModelName{}.

\section{Results}

We evaluate both the baseline models and \IFTModelName{} against security benchmarks described in Section~\ref{sec:sec_benchmarks} and against general evaluations described in Section~\ref{sec:post_training_evals}.
On security benchmarks, we ran $10$ trials at temperature $0.3$ and we report the mean accuracy of the model on each benchmark with one standard deviation in Table~\ref{tab:llm-cyber}. 
On general post-training benchmarks, we stick to standard evaluation settings accepted by the community and implemented by \olmes ~\cite{gu2024olmes}. 
We present these results in Table~\ref{tab:llm-post-training}.
More details are presented in Appendix~\ref{app:gen_benchmarks}.

\subsection{Baselines}

We include the following cybersecurity LLMs as baselines for comparison: \whiterabbitneo-7B (formerly WhiteRabbitNeo) \citep{whiterabbitneo} , \primus-base, and \primus-merged \citep{yu2025primus}.  
We also include comparisons to models from the \llamathree{} \citep{grattafiori2024llama}, \gemma{} \citep{team2025gemma} and \qwen{} families \citep{yang2025qwen3} as well as \gptfourOmini{}\footnote{We use the 07.18.2024 checkpoint for \gptfourOmini{}.} \citep{hurst2024gpt}.  

\subsection{Security Benchmark Performance}

\IFTModelName{} achieves \textbf{state-of-the-art performance} on \ctibench{}-RCM beating larger models including \gptfourOmini{} and \llamathree-70B-Instruct. 
When considering performance error bars (within 1 standard deviation), the model is consistently among the top performers on \ctibench{}-MCQA.
This showcases \IFTModelName{}'s strong performance on CTI tasks, mirroring the performance of \PretrainedModelName{}. 
The model is competitive on \cybermetric{}-500, \secbench{} and \seceval{}.

\subsection{General Post-Training Performance }

\IFTModelName{} is \textbf{best-in-class in instruction following ability and human preferred output} among cybersecurity LLMs. 
On \alpacaleval{}, the next-best cybersecurity model has less than half the win-rate while in \ifeval{} the closest model is 8 points behind. 
The model also outperforms \llamathree{}-8B-Instruct by 11 points and 2 points respectively. 

In grade school math, mathematical reasoning, and coding, \IFTModelName{} rivals peer models and is within 3 percentage points of \llamathree{}-8B-Instruct, demonstraing comparable performance. 
Finally, the performance of \IFTModelName{} with respect to \llamathree-8B-Instruct on \mmlu{} hints at a narrow base of knowledge, while \bbh{} suggests potential for further development in reasoning capabilities.

Overall \IFTModelName{}'s strong instruction following ability sets it up for practical deployment in cybersecurity contexts. 

\begin{table*}[thpb]
  \centering
  \resizebox{\textwidth}{!}{
  \begin{threeparttable}
  \setlength{\tabcolsep}{5pt}  
  \renewcommand{\arraystretch}{1.5}
  \begin{tabular}{l @{\hspace{4pt}} c @{\hspace{4pt}} c @{\hspace{4pt}} c @{\hspace{4pt}} c @{\hspace{4pt}} c @{\hspace{4pt}} c}
    \toprule
    Model
     & \ctibench{}-RCM
     & \ctibench{}-MCQA
     & \ctibench{}-VSP
     & \cybermetric{}-500
     & \secbench{}
     & \seceval{} \\
    \midrule
    \gemma 4B-Instruct 
     & 0.382$\pm$0.006
     & 0.578$\pm$0.001
     & 0.775$\pm$0.001
     & 0.768$\pm$0.002
     & 0.615$\pm$0.002
     & 0.813$\pm$0.001 \\
    \qwen-7B-Instruct 
     & 0.572$\pm$0.008
     & 0.644$\pm$0.001
     & 0.804$\pm$0.001
     & 0.859$\pm$0.002
     & 0.767$\pm$0.003
     & 0.870$\pm$0.001 \\
    \whiterabbitneo-7B 
     & 0.664$\pm$0.007
     & 0.645$\pm$0.006
     & 0.808$\pm$0.003
     & 0.869$\pm$0.007
     & 0.716$\pm$0.004
     & 0.880$\pm$0.003 \\
    \llamathree-8B-Instruct 
     & 0.558$\pm$0.007
     & 0.617$\pm$0.004
     & 0.815$\pm$0.002
     & 0.847$\pm$0.005
     & 0.723$\pm$0.010
     & 0.855$\pm$0.003 \\
    \primus-base 
     & 0.639$\pm$0.006
     & 0.652$\pm$0.003
     & 0.754$\pm$0.008
     & 0.854$\pm$0.008
     & 0.713$\pm$0.005
     & 0.841$\pm$0.006 \\
    \primus-merged 
     & 0.666$\pm$0.007
     & 0.643$\pm$0.004
     & 0.790$\pm$0.003
     & 0.856$\pm$0.004
     & 0.721$\pm$0.009
     & 0.849$\pm$0.002 \\
    \gemma 12B-Instruct 
     & 0.361$\pm$0.007
     & 0.629$\pm$0.001
     &\textbf{0.867$\pm$0.001}
     & 0.865$\pm$0.001
     & 0.749$\pm$0.001
     & 0.868$\pm$0.001 \\
    \midrule
    \gemma 27B-Instruct 
     & 0.549$\pm$0.006
     & 0.653$\pm$0.001
     & 0.821$\pm$0.000
     & 0.870$\pm$0.002
     & 0.780$\pm$0.002
     & 0.896$\pm$0.000 \\
    \llamathree-70B-Instruct 
     & 0.623$\pm$0.005
     &\textbf{0.695$\pm$0.002}
     & 0.840$\pm$0.001
     &\textbf{0.930$\pm$0.005}
     &\textbf{0.839$\pm$0.003}
     &\textbf{0.900$\pm$0.002} \\
    \gptfourOmini{}
     & 0.655$\pm$0.005
     & 0.672$\pm$0.003
     & 0.792$\pm$0.004
     & 0.889$\pm$0.003
     & 0.803$\pm$0.003
     & 0.889$\pm$0.001 \\
    \midrule
    \multirow{2}{*}{\textbf{Foundation-Sec-8B-Instruct}} 
     &\textbf{0.692$\pm$0.005}
     & 0.644$\pm$0.003
     & 0.802$\pm$0.004
     & 0.830$\pm$0.005
     & 0.685$\pm$0.006
     & 0.833$\pm$0.003 \\
    
     & \textbf{($\uparrow$24.03\%)}
     & ($\uparrow$4.40\%)
     & ($\downarrow$-1.67\%)
     & ($\downarrow$-2.01\%)
     & ($\downarrow$-5.21\%)
     & ($\downarrow$-2.50\%) \\
    \bottomrule
  \end{tabular}
  \end{threeparttable}
  }
  \vspace{1ex}
  \caption{Performance on the selected cybersecurity benchmarks (temperature 0.3). Reported performance differences are relative to \llamathree{}-8B-Instruct. For \ctibench{}-RCM, \ctibench{}-MCQA, \cybermetric{}-500, \secbench{}, and \seceval{} we report the average accuracy over 10 trials $\pm$ one standard deviation. For \ctibench{}-VSP we report the average CVSS score (see Appendix~\ref{app:eval_details}) over 10 trials  $\pm$ one standard deviation.}
  \label{tab:llm-cyber}
\end{table*}

\begin{table*}[thpb]
  \centering
  \resizebox{\textwidth}{!}{
  \begin{threeparttable}
   \setlength{\tabcolsep}{2pt} 
  \renewcommand{\arraystretch}{1.25}
  \begin{tabular}{l @{\hspace{8pt}} c @{\hspace{8pt}} c @{\hspace{8pt}} c @{\hspace{8pt}} c @{\hspace{8pt}} c @{\hspace{8pt}} c @{\hspace{8pt}} c}
    \toprule
    \textbf{Model}
     & MMLU
     & BBH
     & GSM8k
     & Alpaca Eval 2
     & MATH
     & IFEval
     & HumanEval \\
    \midrule
    \gemma 4B-Instruct 
     & 0.582
     & 0.709
     & 0.780
     & 42.465
     & 0.604
     & 0.773
     & 0.757 \\
    \qwen-7B-Instruct 
     & 0.717
     & 0.709
     & 0.842
     & 28.983
     & 0.699
     & 0.758
     & 0.927 \\
    \whiterabbitneo-7B 
     & 0.664
     & 0.689
     & 0.828
     & 13.738
     & 0.462
     & 0.721
     & 0.950 \\
    \llamathree-8B-Instruct 
     & 0.679
     & 0.725
     & 0.835
     & 24.477
     & 0.425
     & 0.791
     & 0.864 \\
    \primus-base 
     & 0.650
     & 0.698
     & 0.783
     & 11.454
     & 0.327
     & 0.686
     & 0.856 \\
    \primus-merged 
     & 0.660
     & 0.718
     & 0.805
     & 16.082
     & 0.373
     & 0.738
     & 0.874 \\
    \gemma 12B-Instruct 
     & 0.723
     & 0.820
     & 0.870
     & 52.340
     & 0.732
     & 0.852
     & 0.869 \\

    \midrule
    \gemma 27B-Instruct 
     & 0.767
     &\textbf{0.835}
     & 0.907
     &\textbf{64.044}
     &\textbf{0.758}
     & 0.834
     & 0.893 \\
    \llamathree-70B-Instruct 
     &\textbf{0.810}
     & 0.831
     &\textbf{0.942}
     & 33.140
     & 0.560
     &\textbf{0.878}
     &\textbf{0.951} \\
    \gptfourOmini{} 
     & 0.804
     & 0.777
     & 0.832
     & 52.720
     & 0.679
     & 0.834
     & 0.929 \\

    \midrule
    \multirow{2}{*}{\textbf{Foundation-Sec-8B-Instruct}} 
     & 0.602
     & 0.677
     & 0.817
     & 35.453
     & 0.411
     & 0.811
     & 0.843 \\
    
     & ($\downarrow$-11.31\%)
     & ($\downarrow$-6.57\%)
     & ($\downarrow$-2.09\%)
     & ($\uparrow$44.84\%)
     & ($\downarrow$-3.36\%)
     & ($\uparrow$2.57\%)
     & ($\downarrow$-2.44\%) \\

    \bottomrule
  \end{tabular}
  \end{threeparttable}
  }
  \vspace{1ex}
  \caption{Performance on the selected post-training benchmarks. Reported performance differences are relative to \llamathree{}-8B-Instruct. Evaluation parameters and metrics described in Appendix \ref{app:gen_benchmarks}.}
  \label{tab:llm-post-training}
\end{table*}

\subsection{Persona Adaptation}
Cybersecurity workflows involve distinct practitioner roles such as SOC analysts, red teamers, and threat intelligence specialists, each requiring domain-specific language and task framing. 
To assess the model’s ability to emulate such contexts, we evaluate its persona adaptation capabilities through PersonaGym \citep{samuel2024personagym}---a standardized benchmark for assessing LLMs on multi-turn, persona-driven conversations. 
We evaluate our model on a subset of the PersonaGym test split (50 of the original 200 personas). 
As shown in Table~\ref{tab:personagym-results}, \IFTModelName{} demonstrates strong alignment with assigned personas without requiring dedicated personalization fine-tuning. 
Additional evaluation details are provided in Appendix~\ref{app:personagym}.

\begin{table*}[thpb]
\centering
\resizebox{\textwidth}{!}{
\begin{threeparttable}
\renewcommand{\arraystretch}{1.3}
\begin{tabular}{l @{\hspace{8pt}} c @{\hspace{8pt}} c @{\hspace{8pt}} c @{\hspace{8pt}} c @{\hspace{8pt}} c @{\hspace{8pt}} c}
\toprule
\textbf{Model} & \textbf{\shortstack{Action\\Justification}} & \textbf{\shortstack{Expected\\Actions}} & \textbf{\shortstack{Linguistic\\Habits}} & \textbf{\shortstack{Persona\\Consistency}} & \textbf{\shortstack{Toxicity\\Control}} & \textbf{\shortstack{Persona\\Score}} \\
\midrule
Llama 2-8B & 3.96 & 3.87 & 3.77 & 4.12 & 4.18 & 3.98 \\
GPT 3.5 & 4.31 & 4.28 & 3.63 & 4.70 & 4.96 & 4.38 \\
Llama 2-70B & 4.44 & 4.32 & 3.85 & 4.67 & 4.68 & 4.39 \\
Llama 3-8B & 4.55 & 4.43 & 3.97 & 4.77 & 4.74 & 4.49 \\
Claude 3 Haiku & 2.47 & 4.28 & 3.04 & 3.47 & 4.94 & 3.64 \\
Claude 3.5 Sonnet & 4.52 & 4.37 & 3.98 & 4.81 & 4.88 & 4.51 \\
\midrule
\IFTModelName & 4.4 & 4.6 & 3.95 & 4.95 & 5.0 & 4.58 \\
\bottomrule
\end{tabular}
\end{threeparttable}
}
\vspace{1ex}
\caption{PersonaGym Results. Scores range from 1 (lowest) to 5 (highest). Reported results for all models except ours are taken from \cite{samuel2024personagym}.}
\label{tab:personagym-results}
\end{table*}

\subsection{Comparison to \PretrainedModelName{}} \label{sec:forgetting}

To showcase the retention of cybersecurity knowledge during post-training, we present a head-to-head comparison between \IFTModelName{} and \PretrainedModelName{} on cybersecurity evaluations.  Due to the inability of \PretrainedModelName{} to follow precise instructions, we use 5 shot prompting.
We compare the models on \cybermetric{}-500, \ctibench{}-MCQA, and \secbench{}. 

As noted in \citep{kassianik2025llama}, instruction-tuning affects next-token prediction capabilities of models for few-shot settings, often resulting in verbose responses that do not follow the terse templates given by the few-shot examples. 
This complicates the direct comparison between base and instruction-tuned models using few-shot evaluations.  
This is an issue particularly for \ctibench{}-RCM due to its complexity and short-answer format. 
Therefor we exclude it from this analysis. 

Figure~\ref{fig:comparison_to_base} shows our results. We acknowledge that our measure of latent knowledge retention is limited by the diversity in these evaluations and obscured by each model's ability to follow instructions. 

\begin{figure}[h!]
\centering
\includegraphics[width=0.8\textwidth]{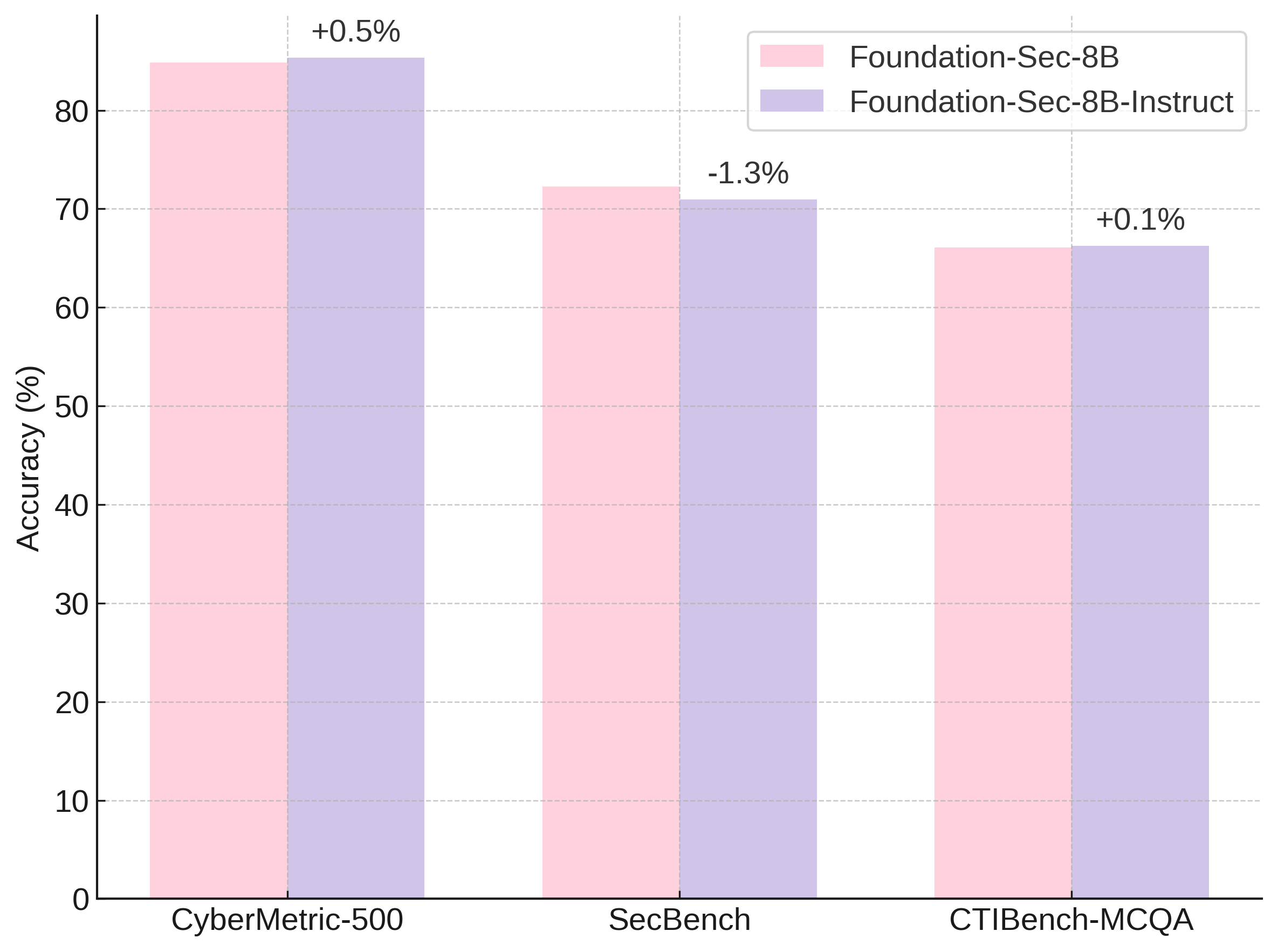}
\caption{Comparison of \IFTModelName{} to \PretrainedModelName{} on selected cybersecurity metrics. The negligible differences in performance suggest minimal forgetting of cybersecurity knowledge in the instruction-tuned model. Both models were 5-shot prompted and results averaged over 10 trials. }
\label{fig:comparison_to_base}
\end{figure} 

\section{Safety Alignment}

\IFTModelName{} has not undergone dedicated safety alignment procedures beyond basic instruction-tuning. However, we followed standard practices to provide a baseline level of alignment. Because this is an instruction-tuned model as opposed to a base model, safety training is particularly important, as the model is designed to follow user prompts and may do so in unsafe ways. 
The model has not been explicitly trained for robustness against adversarial techniques, such as jailbreaking or prompt injection, and no post-hoc safety filtering (e.g., moderation classifiers or rule-based sanitization) has been applied to its outputs. As a result, the model may produce unfiltered or unsafe content, including toxic, biased, or factually incorrect responses. \textbf{We strongly recommend applying additional safety layers, such as automated content filtering or LLM-based moderation systems, when deploying or experimenting with this model.}

To better understand the model’s risk profile and limitations despite the lack of explicit safety alignment, we conducted an evaluation of its safety and robustness using HarmBench \cite{harmbench}, a benchmark framework designed for automated red teaming of LLMs. 
The model was evaluated on 400 representative HarmBench prompts for simple toxicity and demonstrated satisfactory performance, rejecting or safely responding to 92\% of malicious examples across a broad range of risk categories.

To further mitigate safety risks, we recommend pairing the model with LlamaGuard \cite{llamaguard}, which implements taxonomy-driven input-output filtering. When evaluated together, \IFTModelName{} and LlamaGuard rejected 99\% of malicious test cases.

In order to provide users with a good experience when interacting with the model, we developed a detailed system prompt for general user interaction.
We tested this system prompt in internal testing and found that it improved user satisfaction and safety.
This system prompt is baked into the model's chat template but can be modified or overridden as needed.
The system prompt is reproduced in Appendix~\ref{app:system_prompt} and can be tailored to particular use cases.

\begin{figure}[h!]
\centering
\includegraphics[width=0.8\textwidth]{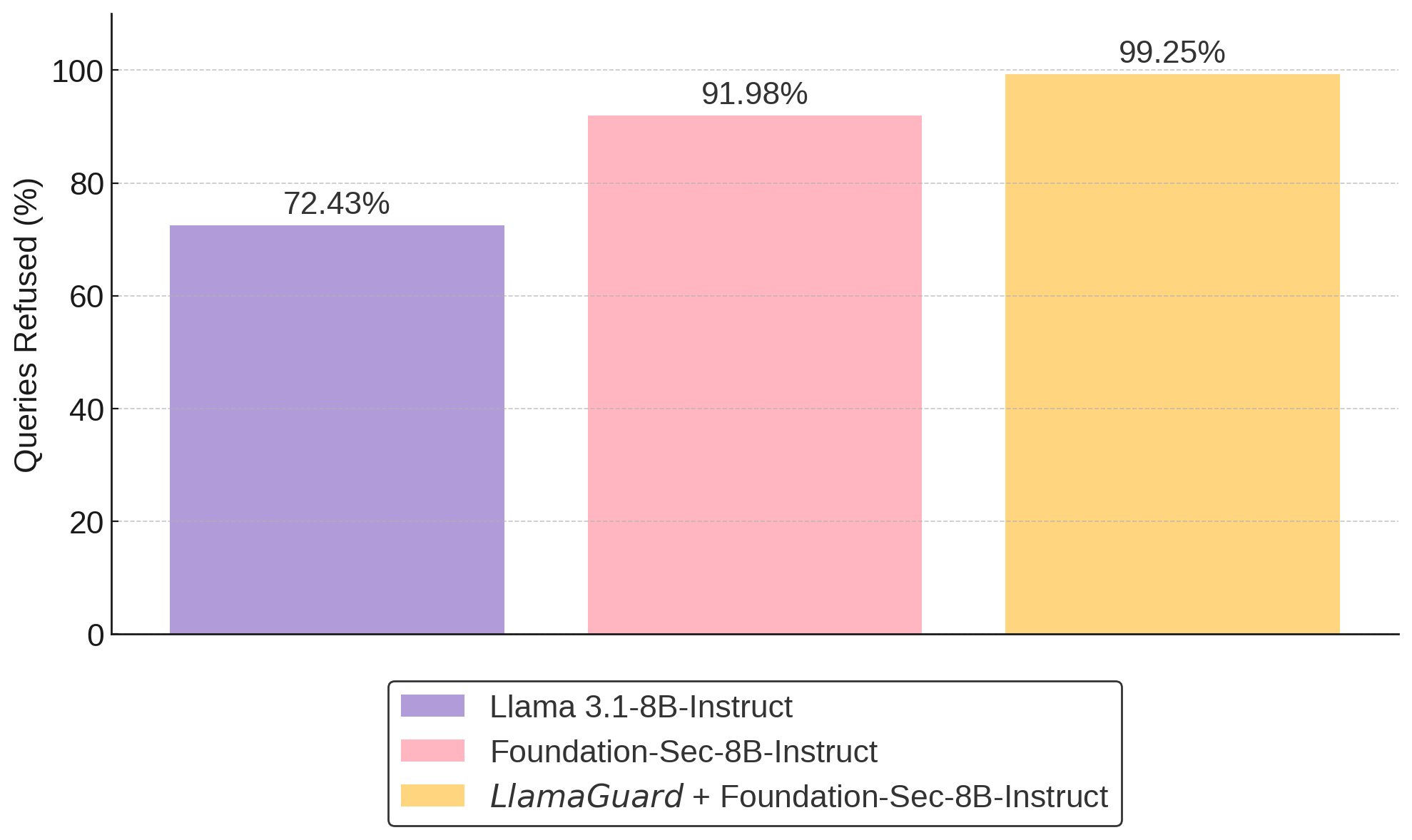}

\caption{Comparison of \llamathree{}-8B-Instruct, \IFTModelName{}, and \IFTModelName{} protected by LlamaGuard \citep{llamaguard} on malicious queries from HarmBench \cite{harmbench}. While \IFTModelName{} performs significantly better than \llamathree{}-8B-Instruct, applying LlamaGuard increases the refusal rate on malicious requests to nearly 100\%. Therefore we recommend deploying the model with additional guardrails in production use cases.}
\end{figure}

\section{Conclusion}

We introduce \IFTModelName{}, a best-in-class large language model (LLM) for general-purpose cybersecurity dialogue, designed to support practitioners, students, and others across a wide range of security tasks. 
\IFTModelName{} outperforms \llamathree{}-8B-Instruct and \gptfourOmini{} on \ctibench{}-RCM and ranks among the top performers in several other cybersecurity benchmarks for its size. 
Beyond domain-specific tasks, it also delivers competitive results on general instruction-following and human preference alignment benchmarks---showcasing its value as a practical model. 
Our results demonstrate that domain-adapted models can retain strong general-purpose capabilities while excelling at specialized tasks. 
With the release of this model, we aim to meet the growing need for cybersecurity-focused LLMs with robust zero-shot performance across diverse contexts. 
We believe \IFTModelName{} will help advance LLM integration in cybersecurity and be a daily driver for many cybersecurity practitioners.

\section*{Acknowledgements}

We thank Alie Fordyce for her leadership in developing a safe model release strategy and extensive documentation. We also thank the Foundation AI team, including Ron Kupfer, Howard Lin, Kimia Majd, Mayank Rajoria, Rahim Dharssi, Alexander Chen, Avi Zohary, Nathan Chang, Roee Landesman, Takahiro Matsumoto, Yasukazu Hirata, Amos Yoffe, Assaf Eisenman, Hyrum Anderson, and Konstantin Goldin for their invaluable support.

\bibliographystyle{plainnat}
\bibliography{bibliography}
\appendix

\section{Security Benchmarks}
\label{app:sec_benchmarks}

\paragraph{\ctibench{} \citep{ctibench}} A benchmark designed to assess LLM performance in cyber threat intelligence (CTI) applications. It targets practical, CTI-specific tasks and includes five sections. We focus on the following: \ctibench{}-MCQA, \ctibench{}-RCM, and \ctibench{}-VSP.

\paragraph{\ctibench{}-Multiple Choice Questions (MCQA)} A 2,500-question benchmark for multiple-choice QA, drawing from CTI frameworks, regulations, and taxonomies such as NIST \citep{nist}, GDPR \cite{gdpr}, MITRE ATT\&CK \citep{mitre_attack}, and CAPEC \citep{capec}.

\paragraph{\ctibench{}-Root Cause Mapping (RCM)} Evaluates a model’s ability to identify the root cause of vulnerabilities by mapping CVE (Common Vulnerability Enumeration) \citep{cve} records and bug reports to CWE (Common Weakness Enumeration) \citep{cwe} entries.

\paragraph{\ctibench{}-Vulnerability Severity Prediction (VSP)} Requires identifying the correct severity of a vulnerability by mapping its description to a Common Vulnerability Scoring System (CVSS) vector string \citep{cvss}. The dataset contains 1,000 samples. While CVSS includes Base, Temporal, and Environmental metric groups, this task focuses on the Base group, which reflects the vulnerability's intrinsic properties. The Base group includes eight metrics: \textit{Attack Vector (AV), Attack Complexity (AC), Privileges Required (PR), User Interaction (UI), Scope (S), Confidentiality Impact (C), Integrity Impact (I), and Availability Impact (A)}. This task demands a nuanced understanding of technical language and context, making it a strong proxy for performance on downstream cybersecurity analytics tasks.

\paragraph{\cybermetric{} \citep{cybermetric}} An MCQA-format dataset created using \gptthreefive{} with Retrieval-Augmented Generation (RAG) applied to a large corpus of security standards, research papers, RFCs, books, and other publications. It was validated through over 200 hours of expert human review to ensure accuracy and relevance. We report results on the 500-sample version of the dataset.

\paragraph{\secbench{} \citep{secbench}} A dataset containing MCQA and short-answer (SAQ) questions designed to assess cybersecurity knowledge and logical reasoning. Unlike \seceval{}, \ctibench{}, and \cybermetric{}, a large portion of \secbench{} questions were sourced from a Cybersecurity Question Design Contest, making the benchmark more relevant and challenging. We use only the multiple-choice questions from the English subset (595 questions).

\paragraph{\seceval{} \citep{li2023seceval}} An MCQA dataset with over 2,000 samples spanning nine cybersecurity domains: Software Security, Application Security, System Security, Web Security, Cryptography, Memory Safety, Network Security, and PenTesting.

\paragraph{\mmlusec{} \citep{mmlu}} The computer security subset of the MMLU (Measuring Massive Multitask Language Understanding) benchmark \cite{mmlu}. It includes 116 questions covering topics such as cryptography, malware, and fuzzing.

\subsection{Evaluation Details}
\label{app:eval_details}

For evaluations on \ctibench{}-MCQA, \ctibench{}-RCM, \cybermetric{}, \secbench{}, and \seceval{}, we follow the same approach used in our base model’s development. For more details, see \citep[Appendix B.3]{kassianik2025llama}.

\paragraph{\ctibench{}-VSP}

Models are tasked with predicting the CVSS v3.1 vector for a given CVE description:

\begin{tcolorbox}[
    title=Linux Kernel Vulnerability,
    colback=white,
    colframe=black,
    fonttitle=\bfseries,
    breakable,
    left=5pt,
    right=5pt,
    top=5pt,
    bottom=5pt
]
\begin{center}
\begin{minipage}{0.95\textwidth}
\begin{lstlisting}[
    frame=none,
    framesep=5pt,
    xleftmargin=5pt,
    xrightmargin=5pt,
    basicstyle=\small\ttfamily\linespread{1.2}\selectfont,
    breaklines=true,
    breakautoindent=false,
    breakindent=0pt,
    numbers=none
]
In the Linux kernel through 6.7.1, there is a use-after-free in cec_queue_msg_fh, related to drivers/media/cec/core/cec-adap.c and drivers/media/cec/core/cec-api.c.
\end{lstlisting}
\end{minipage}
\end{center}
\end{tcolorbox}

We guide all models to compute the CVSS vector using the following instruction, which is a slightly modified version of the original benchmark prompt:

\begin{tcolorbox}[
    title=CVSS Instructions,
    colback=white,
    colframe=black,
    fonttitle=\bfseries,
    breakable,
    left=5pt,
    right=5pt,
    top=5pt,
    bottom=5pt
]
\begin{center}
\begin{minipage}{0.95\textwidth}
\begin{lstlisting}[
    frame=none,
    framesep=2pt,
    xleftmargin=2pt,
    xrightmargin=2pt,
    basicstyle=\small\ttfamily\linespread{1.2}\selectfont,
    breaklines=true,
    breakautoindent=false,
    breakindent=0pt,
    numbers=none
]
From the following CVE description, determine the CVSS v3.1 vector string for each CVSS base metric: AV, AC, PR, UI, S, C, I, and A.
Valid options for each metric are as follows:
  - **Attack Vector (AV)**: Network (N), Adjacent (A), Local (L), Physical (P)
  - **Attack Complexity (AC)**: Low (L), High (H)
  - **Privileges Required (PR)**: None (N), Low (L), High (H)
  - **User Interaction (UI)**: None (N), Required (R)
  - **Scope (S)**: Unchanged (U), Changed (C)
  - **Confidentiality (C)**: None (N), Low (L), High (H)
  - **Integrity (I)**: None (N), Low (L), High (H)
  - **Availability (A)**: None (N), Low (L), High (H)
Provide your answer as a CVSS v3.1 vector string. Ensure your final answer contains only the CVSS v3 Vector String in the following example format:
CVSS:3.1/AV:N/AC:L/PR:N/UI:N/S:U/C:H/I:H/A:H
\end{lstlisting}
\end{minipage}
\end{center}
\end{tcolorbox}

Each predicted CVSS vector is converted to a CVSS base score between 0.0 and 10.0, representing the overall threat level. For the VSP benchmark, we use a metric derived from the mean absolute deviation (MAD) of the base score, following \citet{ctibench}, but normalize and invert it so that a perfect match yields a score of 1.0:
\[
  \operatorname{CVSS\_Score} = 1 - \frac{\operatorname{MAD}}{10}
\]
We report the $\mathrm{CVSS\_Score}$ to simplify comparisons across benchmarks.

In contrast to the original \ctibench{}-VSP setup, which excluded malformed predictions that could not be parsed as valid CVSS vectors, we assign them the maximum deviation of $10.0$. Lastly, we set a higher $\mathrm{max\_tokens} = 1000$, as models often explain their choices for each component before outputting the final CVSS vector.

\begin{tcolorbox}[
    title=MCQA prompt format,
    colback=white,
    colframe=black,
    fonttitle=\bfseries,
    breakable,
    left=5pt,
    right=5pt,
    top=5pt,
    bottom=5pt
]
\begin{center}
\begin{minipage}{0.95\textwidth}
\begin{lstlisting}[
    frame=none,
    framesep=2pt,
    xleftmargin=2pt,
    xrightmargin=2pt,
    basicstyle=\small\ttfamily\linespread{1.2}\selectfont,
    breaklines=true,
    breakautoindent=false,
    breakindent=0pt,
    numbers=none
]
Given the following question and four candidate answers (A, B, C and D), choose the best answer. Your response should be of the following format: 'Answer: \$LETTER' (without quotes) where LETTER is one of A, B, C, or D.
Three of the following are classic security properties; which one is not?
A. Confidentiality
B. Availability
C. Correctness
D. Integrity
\end{lstlisting}
\end{minipage}
\end{center}
\end{tcolorbox}
\captionof{figure}{Zero-shot prompt format used with instruct-finetuned models for MCQA tasks.}
\label{fig:ift_input_tmp_mcqa}

\begin{tcolorbox}[
    title=CWE ID prompt format,
    colback=white,
    colframe=black,
    fonttitle=\bfseries,
    breakable,
    left=5pt,
    right=5pt,
    top=5pt,
    bottom=5pt
]
\begin{center}
\begin{minipage}{0.95\textwidth}
\begin{lstlisting}[
    frame=none,
    framesep=2pt,
    xleftmargin=2pt,
    xrightmargin=2pt,
    basicstyle=\small\ttfamily\linespread{1.2}\selectfont,
    breaklines=true,
    breakautoindent=false,
    breakindent=0pt,
    numbers=none
]
Analyze the following CVE description and map it to the appropriate CWE. Provide a brief justification for your choice. Ensure the last line of your response contains only the CWE ID.
CVE Description: Tenda AX1803 v1.0.0.1 contains a stack overflow via the iptv.city.vlan parameter in the function getIptvInfo.
\end{lstlisting}
\end{minipage}
\end{center}
\end{tcolorbox}
\captionof{figure}{Zero-shot prompt format used with instruct-finetuned models for the CWE ID mapping task, i.e., \ctibench{}-RCM.}
\label{fig:ift_input_tmp_cwe}

\section{General Capability Benchmarks} \label{app:gen_benchmarks}

\paragraph{\mmlu{} \citep{mmlu}} A diverse MCQA benchmark spanning 57 subjects, including the humanities, STEM, law, and history. Its broad coverage allows us to assess the knowledge retained after training \IFTModelName{}, helping us detect signs of overfitting or catastrophic forgetting.

\paragraph{\humaneval{} \citep{chen2021evaluating}} Measures Python programming ability using prompts with docstrings. Programming is a critical skill in the cybersecurity field, and many cybersecurity research tasks involve programming. This benchmark helps gauge how well \IFTModelName{} retains such capabilities.

\paragraph{\bbh{} \citep{suzgun2022challenging}} A challenging subset of \bigbench{} \citep{srivastava2023beyond}, where prior LLMs have underperformed compared to average human raters. Success on \bbh{} requires strong reasoning across arithmetic, logic, geometry, spatial and temporal tasks, as well as general and semantic knowledge. We use it to evaluate the model’s underlying reasoning ability.

\paragraph{\ifeval{} \citep{zhou2023instruction}} Tests a model’s ability to follow verifiable instructions (e.g., “output as a JSON object”). It includes 25 instruction types across roughly 500 prompts. Strong performance on \ifeval{} is indicative of downstream usability in cybersecurity workflows.

\paragraph{\alpacaleval{} \citep{dubois2024length}} An automatic evaluation metric that scores a model’s chat responses based on alignment with human preferences. Unlike other benchmarks, it is reference-free—open-ended questions with no ground-truth answers. We use \alpacaleval{} to assess how well the model performs as a chatbot and how closely its responses align with user expectations.

\paragraph{\gsmk{} \citep{cobbe2021gsm8k}} A dataset of high-quality grade school math problems involving moderate difficulty and early algebra. While not directly related to cybersecurity, the domain’s broad demands justify evaluating on \gsmk{} to build a more complete view of overall model performance.

\paragraph{\matheval{} \citep{hendrycks2021measuring}} A collection of 12,500 math competition problems requiring advanced reasoning beyond grade school level. Solving them demands multi-step thinking and heuristics. We include \matheval{} to probe \IFTModelName{}’s ability for structured, step-by-step reasoning—key for tackling complex cybersecurity challenges.

\begin{table}[h]
\centering
    \begin{tabular}{lccc}
    \toprule
    \textbf{Benchmark} & \textbf{Metric} & \textbf{Notes} \\
    \midrule
    \mmlu{} & exact match & weighted macro average \\
    \humaneval{} & pass@10 & temperature 0.8 \\
    \bbh{} & exact match & 3-shot CoT  \\
    \ifeval{} & pass@1  & prompt loose average \\
    \alpacaleval{} & length controlled win rate & 0-shot \\
    \gsmk{} & exact match & 8-shot CoT  \\
    \matheval{} & flex exact match & 4-shot CoT \\
    \bottomrule
    \end{tabular}
\caption{Summary of the post-training evaluation metrics. For \matheval{}, we adopt the `flex, exact match` evaluation scheme from \cite{tulu3}.}
\label{tab:post_training_metrics}
\end{table}

\section{Cybersecurity Taxonomy} \label{app:taxonomy}

We organize the topic of cybersecurity into the following categories:

\begin{enumerate}
    \item \textbf{Governance, Risk, and Compliance (GRC)}
    \begin{itemize}
        \item Risk Management \& Security Strategy
        \item Compliance and Regulations (e.g., GDPR, HIPAA)
        \item Security Frameworks (e.g., NIST CSF, ISO 27001)
        \item Security Policies \& Architecture
    \end{itemize}
    
    \item \textbf{Network, Infrastructure, and Endpoint Security}
    \begin{itemize}
        \item Perimeter and Network Security (Firewalls, VPNs, Wireless)
        \item Endpoint Protection \& MDM
        \item IoT and OT/ICS Security
        \item Mobile Security
    \end{itemize}
    
    \item \textbf{Application and Software Security}
    \begin{itemize}
        \item Secure Software Development (DevSecOps)
        \item Application \& API Security
        \item Vulnerability Management \& Penetration Testing
        \item Software Supply Chain Security (SBOM, third-party risk)
    \end{itemize}
    
    \item \textbf{Cloud and Data Security}
    \begin{itemize}
        \item Cloud Security Architecture \& Tools
        \item Identity and Access Management (IAM, PAM)
        \item Data Loss Prevention \& Privacy (DLP, encryption)
        \item Cloud Compliance \& Shared Responsibility Model
    \end{itemize}
    
    \item \textbf{Identity, Access, and Zero Trust}
    \begin{itemize}
        \item Authentication \& Authorization (MFA, SSO, RBAC)
        \item Identity Governance \& Lifecycle
        \item Zero Trust Architecture
        \item Privileged Access Controls
    \end{itemize}
    
    \item \textbf{Security Operations and Monitoring (SecOps)}
    \begin{itemize}
        \item SIEM, SOC, and Log Management
        \item Security Automation \& SOAR
        \item Detection Engineering
        \item Operational Resilience \& Monitoring
    \end{itemize}
    
    \item \textbf{Threat Intelligence and Incident Response}
    \begin{itemize}
        \item Threat Detection, Analysis \& Hunting
        \item Threat Intelligence Platforms \& IOCs
        \item Advanced Persistent Threats (APTs)
        \item Malware Techniques
        \item Incident Response, Recovery \& Digital Forensics
    \end{itemize}
    
    \item \textbf{Cryptography and Secure Communications}
    \begin{itemize}
        \item Cryptographic Algorithms \& PKI
        \item Key Management
        \item Post-Quantum Cryptography
        \item Secure Protocols and Encryption Practices
    \end{itemize}
    
    \item \textbf{Security Awareness and Human Risk}
    \begin{itemize}
        \item Social Engineering Techniques (Phishing, Pretexting)
        \item Insider Threat Management
        \item Security Awareness Training
        \item Behavioral Risk Analysis
    \end{itemize}
    
    \item \textbf{Emerging Technologies and Future Threats}
    \begin{itemize}
        \item AI/ML \& LLM Security (adversarial ML, prompt injection)
        \item Quantum Security Threats
        \item Deepfakes \& Synthetic Media
        \item Nation-State Threats and Geopolitical Risk
    \end{itemize}
\end{enumerate}

\section{Post-Training Data Analysis}\label{app:post_training_data_analysis}

Here we present further details into our estimates for cybersecurity content in open-sourced post-training datasets. We limit our analysis to only prompts and do not consider responses.

\subsection{Keyword Filtering Methodology}
 
We apply a series of pattern matching filters to detect cybersecurity relevance. The set of 795 keywords that is used in this study draws from general cybersecurity concepts (e.g., \texttt{authentication}, \texttt{encryption}, and \texttt{vulnerability}), cybersecurity tools and frameworks (e.g., \texttt{SIEM} and \texttt{SOAR}), threat landscape terminology (e.g., \texttt{zero-day exploit}), and types of attacks (e.g., \texttt{SQL injection} and \texttt{cross-site scripting}). In addition, we used regex patterns to match various structured technical identifiers, including CVEs, VWEs, MITRE ATT\&CK Techniques, and NIST Control Identifiers.
Data is preprocessed prior to being sent to the filters in a series of steps, including URL and email removal, special character and whitespace normalization. All the filtering we do is case-insensitive.

After initial passes through the datasets, we discovered two main issues that were adversely affecting our classification accuracy. The first was that non-English texts were not being matched by our filters, since our keyword bank only covered English terms, which increased our false negative rate. The second issue was that a handful of keywords were contributing to substantial false positives, because they were too general and had additional meanings extending beyond cybersecurity. These included common words like \texttt{``sandbox''}, \texttt{``policy''}, and \texttt{``secure''}, and abbreviations like \texttt{``APT''}, \texttt{``DOS''}, and \texttt{``C2''}.

We addressed both of these issues in our final evaluation of the datasets. 
We used a FastText language detection model \citep{joulin2016bag} to filter out non-English prompts from our analysis, to avoid underestimating the true percentages of cybersecurity-relevant prompts. 
We also excluded the 17 identified keywords that incur high false positive rates from the original list of 795 when computing if the number of matches exceeds the threshold for classification. 
Note that the Average Matches and Average Density statistics in Table~\ref{tab:app_keyword_estimate_detailed} do still include these keywords. 

The filters are parameterized by a keyword threshold. We balanced the threshold based on careful tuning, so it can work in tandem with the cybersecurity classification model in the second part of the pipeline.

\begin{table}[h]
\centering
\resizebox{\textwidth}{!}{
\begin{tabular}{lccccc}
\toprule
\textbf{Source} & \textbf{Hits} & \textbf{\% Dataset} & \textbf{Average Matches} & \textbf{Average Density} & \textbf{Dataset Size} \\
\midrule
\openorca{} & 37,451 & 0.88\% & 1.4 & 0.8\% & 4,233,923 \\
\tuluthree{} SFT Mixture & 17,264 & 1.84\% & 2.0 & 2.4\% & 939,343 \\
\tulutwo{} SFT Mixture & 8,021 & 2.46\% & 1.8 & 1.32\% & 326,154 \\
\wizardevol{} & 1,479 & 2.11\% & 2.2 & 2.7\% & 70,000 \\
\nemotronchat{} & 806 & 2.03\% & 1.8 & 3.46\% & 39,792 \\
\nemotronsafety{} & 216 & 0.69\% & 1.6 & 2.03\% & 31,426 \\
\openplatypus{} & 184 & 0.74\% & 1.8 & 1.63\% & 24,926 \\
\alpaca{} & 172 & 0.33\% & 1.2 & 14.43\% & 52,002 \\
\bottomrule
\end{tabular}
}
\caption{Keyword-based estimates of cybersecurity content in post-training datasets. Keyword threshold: 1.  Average Matches and Average Density are only computed for prompts above the keyword threshold.}
\label{tab:app_keyword_estimate_detailed}
\end{table}

\clearpage

\subsection{Examples of Cybersecurity Content Identified}

\begin{tcolorbox}[
    title=Example Cybersecurity Training Sample from \tuluthree{},
    colback=white,
    colframe=black,
    fonttitle=\bfseries,
    breakable
]

\textbf{Source:} tulu-3-sft-mixture

\textbf{ID:} ai2-adapt-dev/evol\_codealpaca\_heval\_decontaminated\_93564

\textbf{Keywords Found:} \colorbox{blue!20}{hashing}, \colorbox{blue!20}{password strength}, \colorbox{blue!20}{salting}, \colorbox{blue!20}{security}

\vspace{6pt}
\textbf{Training Sample Content:}

\begingroup
\setlength{\parskip}{4pt}
\ttfamily\small

Develop a Java program that utilizes the `sha1` \colorbox{blue!20}{hashing} algorithm along with a salt for additional \colorbox{blue!20}{security} to encrypt a user's password. The program must then store this encrypted password securely.

For an added level of complexity, your program must handle user input for the password and the salt must be randomly generated each time the program runs for the same user's password. The salt alongside the hashed password should then be saved such that the original password can still be verified but cannot be retrieved directly.

The program should also prevent users from choosing weak passwords such as short lengths, commonly used words or sequences, etc. Implement a \colorbox{blue!20}{password strength} checker, that enforces complex password rules such as a minimum length, inclusion of uppercase and lowercase letters, numbers, and special characters.

Also handle any exceptions that may occur during the execution and document the code properly for better readability and future enhancements.

Test your program with multiple inputs to ensure that \colorbox{blue!20}{salting} and \colorbox{blue!20}{hashing} are working correctly, and that weak passwords are rejected.

Please also provide an example of how to verify a password input against the stored, hashed password.

Hints: You will need to use a library such as 'javax.crypto' for creating the hashed password, and Java's SecureRandom class to generate random numbers for the salt.
\endgroup
\end{tcolorbox}

\begin{tcolorbox}[
    title=Example Cybersecurity Training Sample from Nemotron,
    colback=white,
    colframe=black,
    fonttitle=\bfseries,
    breakable
]

\textbf{Source:} nemotron-sft-chat

\textbf{ID:} 35961

\textbf{Keywords Found:} \colorbox{blue!20}{decryption}, \colorbox{blue!20}{mitre}, \colorbox{blue!20}{nist}, \colorbox{blue!20}{security}

\vspace{6pt}
\textbf{Training Sample Content:}

\begingroup
\setlength{\parskip}{4pt}
\ttfamily\small
Write a 200-word free verse poem that personifies various \colorbox{blue!20}{security} controls as diligent guardians, emphasizing their role in protecting digital assets. The poem should explore the concept of \colorbox{blue!20}{security} through the metaphor of a well-maintained fortress, highlighting the importance of regular audits in upholding the strength and resilience of these cybersecurity defenses

--- ASSISTANT MESSAGE ---

How would you modify the poem to address the challenges of zero-day exploits and AI-driven attacks, while maintaining the fortress metaphor and incorporating modern hybrid cloud environments?

--- ASSISTANT MESSAGE ---

Can you revise the poem to integrate real-world cybersecurity frameworks like \colorbox{blue!20}{MITRE} ATT\&CK or \colorbox{blue!20}{NIST} CSF as architectural blueprints, while contrasting legacy systems as crumbling wing of the fortress vulnerable to quantum \colorbox{blue!20}{decryption}?
\endgroup
\end{tcolorbox}

\section{Decontamination}
\label{app:decontamination}

\subsection{Methodology}
\label{app:decont_method}

To mitigate evaluation leakage, we implement a layered decontamination methodology aimed at capturing a wide spectrum of contamination in post-training corpora. 
Such contamination, whether exact duplicates, partial matches, or semantically similar paraphrases, can artificially inflate benchmark performance and mask a model’s true generalization ability. 
Our framework addresses this challenge through two complementary stages: surface-level detection and semantic similarity retrieval.

\begin{table*}
\tiny  
\centering
\setlength{\tabcolsep}{1.5pt}  
\renewcommand{\arraystretch}{1.6}
\definecolor{custommustard}{HTML}{ffecec}
\definecolor{customred}{HTML}{ffcccc}
\definecolor{darkred}{HTML}{ff9797}
\begin{tabularx}{\textwidth}{l*{20}{>{\centering\arraybackslash}X}}  
\toprule
 & \multicolumn{4}{c}{\textbf{Tulu 3 SFT}} & \multicolumn{4}{c}{\textbf{Tulu 2 SFT}} & \multicolumn{4}{c}{\textbf{Open-Platypus}} & \multicolumn{4}{c}{\textbf{Nemotron-Safety}} & \multicolumn{4}{c}{\textbf{Nemotron-Chat}} \\
\cmidrule(lr){2-5} \cmidrule(lr){6-9} \cmidrule(lr){10-13} \cmidrule(lr){14-17} \cmidrule(lr){18-21}
Benchmark & EM & NG & ES & \mbox{LLM-ES} & EM & NG & ES & \mbox{LLM-ES} & EM & NG & ES & \mbox{LLM-ES} & EM & NG & ES & \mbox{LLM-ES} & EM & NG & ES & \mbox{LLM-ES} \\
\midrule
\multicolumn{21}{c}{\textit{Post-Training Benchmarks}} \\
\midrule
BigBenchHard  
 & \cellcolor{custommustard}0.02 & \cellcolor{custommustard}0.23 & \cellcolor{customred}14.52 & \cellcolor{custommustard}0.3 
 & \cellcolor{custommustard}0.02 & \cellcolor{custommustard}0.11 & \cellcolor{customred}4.13 & \cellcolor{custommustard}0.1 
 & 0.00 & \cellcolor{custommustard}0.02 & \cellcolor{customred}8.57 & \cellcolor{custommustard}0.1 
 & 0.00 & \cellcolor{custommustard}0.02 & 0.00 & 0.0 
 & 0.00 & \cellcolor{custommustard}0.02 & \cellcolor{custommustard}0.46 & 0.0 \\
GSM8K  
 & 0.00 & \cellcolor{custommustard}0.15 & \cellcolor{darkred}82.99 & \cellcolor{customred}4.4 
 & 0.00 & \cellcolor{custommustard}0.08 & \cellcolor{darkred}69.83 & \cellcolor{custommustard}0.6 
 & 0.00 & 0.00 & \cellcolor{darkred}63.53 & \cellcolor{custommustard}0.1 
 & 0.00 & 0.00 & 0.00 & 0.0 
 & 0.00 & 0.00 & \cellcolor{custommustard}0.68 & \cellcolor{custommustard}0.1 \\
HumanEval  
 & 0.00 & 0.00 & \cellcolor{darkred}100.00 & \cellcolor{customred}46.3 
 & 0.00 & 0.00 & \cellcolor{darkred}76.22 & \cellcolor{customred}23.8 
 & 0.00 & 0.00 & \cellcolor{darkred}57.32 & \cellcolor{custommustard}1.8 
 & 0.00 & 0.00 & 0.00 & 0.0 
 & 0.00 & 0.00 & \cellcolor{customred}4.27 & \cellcolor{custommustard}1.2 \\
IFEval  
 & 0.00 & 0.00 & \cellcolor{customred}28.83 & \cellcolor{customred}2.8 
 & 0.00 & 0.00 & \cellcolor{customred}21.07 & \cellcolor{custommustard}0.7 
 & 0.00 & 0.00 & \cellcolor{customred}2.40 & 0.0 
 & 0.00 & 0.00 & \cellcolor{customred}2.22 & \cellcolor{custommustard}0.2 
 & 0.00 & 0.00 & \cellcolor{customred}4.25 & 0.0 \\
MATH  
 & 0.00 & \cellcolor{custommustard}0.30 & \cellcolor{darkred}52.58 & \cellcolor{customred}5.3 
 & 0.00 & 0.00 & \cellcolor{customred}9.44 & 0.0 
 & \cellcolor{darkred}97.80 & \cellcolor{customred}42.78 & \cellcolor{darkred}99.78 & \cellcolor{darkred}91.0 
 & 0.00 & 0.00 & 0.00 & 0.0 
 & 0.00 & 0.00 & \cellcolor{custommustard}0.72 & \cellcolor{custommustard}0.0 \\
MMLU  
 & \cellcolor{custommustard}0.96 & \cellcolor{custommustard}0.75 & \cellcolor{customred}21.14 & \cellcolor{customred}4.4 
 & \cellcolor{custommustard}0.88 & \cellcolor{custommustard}0.38 & \cellcolor{customred}14.41 & \cellcolor{custommustard}0.5 
 & \cellcolor{custommustard}0.21 & \cellcolor{custommustard}0.16 & \cellcolor{customred}11.64 & \cellcolor{custommustard}0.3 
 & \cellcolor{custommustard}0.01 & 0.00 & \cellcolor{custommustard}1.18 & \cellcolor{custommustard}0.0 
 & \cellcolor{custommustard}0.18 & \cellcolor{custommustard}0.04 & \cellcolor{customred}2.17 & \cellcolor{custommustard}0.1 \\
AlpacaEval  
 & \cellcolor{custommustard}0.87 & \cellcolor{customred}2.61 & \cellcolor{customred}40.67 & \cellcolor{customred}8.9 
 & \cellcolor{customred}4.60 & \cellcolor{customred}22.73 & \cellcolor{darkred}50.08 & \cellcolor{customred}27.0 
 & \cellcolor{customred}5.09 & \cellcolor{customred}4.35 & \cellcolor{customred}10.43 & \cellcolor{customred}5.6 
 & 0.00 & 0.00 & \cellcolor{customred}2.61 & \cellcolor{custommustard}0.6 
 & \cellcolor{custommustard}0.12 & \cellcolor{custommustard}0.25 & \cellcolor{customred}6.21 & \cellcolor{custommustard}0.8 \\
\midrule
\multicolumn{21}{c}{\textit{Security Benchmarks}} \\
\midrule
CTI-ATE  
 & 0.00 & 0.00 & 0.00 & 0.0 
 & 0.00 & 0.00 & \cellcolor{customred}3.33 & 0.0 
 & 0.00 & 0.00 & 0.00 & 0.0 
 & 0.00 & 0.00 & 0.00 & 0.0 
 & 0.00 & 0.00 & 0.00 & 0.0 \\
CTI-MCQA  
 & 0.00 & 0.00 & \cellcolor{customred}4.12 & \cellcolor{custommustard}0.1 
 & 0.00 & \cellcolor{custommustard}0.08 & \cellcolor{customred}3.44 & \cellcolor{custommustard}0.0 
 & 0.00 & 0.00 & \cellcolor{custommustard}0.08 & 0.0 
 & 0.00 & 0.00 & \cellcolor{custommustard}0.04 & 0.0 
 & 0.00 & 0.00 & \cellcolor{custommustard}0.24 & 0.0 \\
CTI-RCM  
 & 0.00 & 0.00 & \cellcolor{custommustard}1.70 & 0.0 
 & 0.00 & 0.00 & \cellcolor{customred}3.70 & \cellcolor{custommustard}0.3 
 & 0.00 & 0.00 & 0.00 & 0.0 
 & 0.00 & 0.00 & 0.00 & 0.0 
 & 0.00 & 0.00 & \cellcolor{custommustard}0.10 & 0.0 \\
CTI-TAA  
 & 0.00 & 0.00 & \cellcolor{customred}42.00 & \cellcolor{customred}2.0 
 & 0.00 & 0.00 & \cellcolor{customred}16.00 & 0.0 
 & 0.00 & 0.00 & 0.00 & 0.0 
 & 0.00 & 0.00 & \cellcolor{customred}2.00 & 0.0 
 & 0.00 & 0.00 & \cellcolor{customred}2.00 & 0.0 \\
CTI-VSP  
 & 0.00 & 0.00 & \cellcolor{customred}3.70 & 0.0 
 & 0.00 & 0.00 & \cellcolor{customred}2.90 & 0.0 
 & 0.00 & 0.00 & 0.00 & 0.0 
 & 0.00 & 0.00 & 0.00 & 0.0 
 & 0.00 & 0.00 & 0.00 & 0.0 \\
SecBench  
 & 0.00 & \cellcolor{custommustard}0.34 & \cellcolor{customred}23.87 & \cellcolor{custommustard}0.2 
 & 0.00 & \cellcolor{custommustard}0.34 & \cellcolor{customred}32.61 & 0.0 
 & 0.00 & 0.00 & \cellcolor{custommustard}0.84 & 0.0 
 & 0.00 & 0.00 & 0.00 & 0.0 
 & 0.00 & 0.00 & \cellcolor{customred}4.71 & \cellcolor{custommustard}0.2 \\
SecBench-R  
 & 0.00 & 0.00 & \cellcolor{customred}16.28 & 0.0 
 & 0.00 & 0.00 & \cellcolor{customred}27.91 & 0.0 
 & 0.00 & 0.00 & 0.00 & 0.0 
 & 0.00 & 0.00 & 0.00 & 0.0 
 & 0.00 & 0.00 & 0.00 & 0.0 \\
MMLU-Sec  
 & 0.00 & 0.00 & \cellcolor{customred}17.00 & 0.0 
 & 0.00 & 0.00 & \cellcolor{customred}20.00 & \cellcolor{custommustard}1.0 
 & 0.00 & 0.00 & \cellcolor{customred}2.00 & 0.0 
 & 0.00 & 0.00 & \cellcolor{custommustard}1.00 & 0.0 
 & 0.00 & 0.00 & \cellcolor{custommustard}1.00 & 0.0 \\
SecEval  
 & 0.00 & 0.00 & \cellcolor{customred}14.98 & \cellcolor{custommustard}0.6 
 & 0.00 & 0.00 & \cellcolor{customred}11.08 & \cellcolor{custommustard}0.6 
 & 0.00 & 0.00 & \cellcolor{custommustard}0.48 & \cellcolor{custommustard}0.1 
 & 0.00 & 0.00 & \cellcolor{custommustard}0.08 & 0.0 
 & 0.00 & 0.00 & \cellcolor{custommustard}0.96 & 0.0 \\
CM-10K  
 & \cellcolor{custommustard}0.04 & \cellcolor{custommustard}0.23 & \cellcolor{customred}23.93 & \cellcolor{custommustard}1.1 
 & \cellcolor{custommustard}0.11 & \cellcolor{custommustard}0.43 & \cellcolor{customred}25.40 & \cellcolor{custommustard}1.8 
 & 0.00 & 0.00 & \cellcolor{customred}2.36 & \cellcolor{custommustard}0.0 
 & 0.00 & 0.00 & \cellcolor{custommustard}0.64 & \cellcolor{custommustard}0.0 
 & \cellcolor{custommustard}0.06 & \cellcolor{custommustard}0.07 & \cellcolor{customred}2.44 & \cellcolor{custommustard}0.3 \\
CM-2K  
 & \cellcolor{custommustard}0.05 & \cellcolor{custommustard}0.10 & \cellcolor{customred}23.80 & \cellcolor{custommustard}0.8 
 & \cellcolor{custommustard}0.10 & \cellcolor{custommustard}0.20 & \cellcolor{customred}25.05 & \cellcolor{custommustard}1.9 
 & 0.00 & 0.00 & \cellcolor{customred}2.30 & 0.0 
 & 0.00 & 0.00 & \cellcolor{custommustard}0.45 & \cellcolor{custommustard}0.1 
 & \cellcolor{custommustard}0.10 & \cellcolor{custommustard}0.05 & \cellcolor{customred}2.05 & \cellcolor{custommustard}0.1 \\
CM-500  
 & 0.00 & 0.00 & \cellcolor{customred}28.00 & \cellcolor{custommustard}0.6 
 & 0.00 & \cellcolor{custommustard}0.20 & \cellcolor{customred}25.00 & \cellcolor{custommustard}1.6 
 & 0.00 & 0.00 & \cellcolor{customred}3.20 & 0.0 
 & 0.00 & 0.00 & \cellcolor{custommustard}1.20 & 0.0 
 & 0.00 & 0.00 & \cellcolor{customred}2.80 & \cellcolor{custommustard}0.2 \\
CM-80  
 & 0.00 & 0.00 & \cellcolor{customred}35.00 & \cellcolor{customred}2.5 
 & 0.00 & \cellcolor{custommustard}1.25 & \cellcolor{customred}28.75 & \cellcolor{custommustard}1.3 
 & 0.00 & 0.00 & \cellcolor{customred}3.75 & 0.0 
 & 0.00 & 0.00 & 0.00 & 0.0 
 & 0.00 & 0.00 & \cellcolor{customred}2.50 & 0.0 \\
 \bottomrule
\end{tabularx}
\caption{Detailed data contamination analysis across different datasets and benchmarks (Part 1). We report percentages of counts with respect to benchmarks rounded off (\%). EM = Exact Match, NG = N-Gram Overlap, ES = Embedding Similarity, LLM-ES = LLM-based Embedding Similarity. CTI = \ctibench{}, CM = \cybermetric{}, SecBench-R = \secbench{}-Reasoning, MMLU-Sec = MMLU-Security. \colorbox{custommustard}{Light red} indicates values $<2\%$ but $>0\%$, \colorbox{customred}{red} indicates values $\ge2\%$ but $<50\%$, \colorbox{darkred}{dark red} indicates values $\ge50\%$.}
\label{tab:detailed_contamination_part1}
\end{table*}

\paragraph{Stage 1: Surface-Level Overlap}
We begin with exact match detection to identify benchmark samples that appear verbatim in the training corpus, capturing the most direct forms of leakage. To detect substantial partial matches, we also perform n-gram overlap analysis using a sliding 8‑gram window. A sample is flagged if over 50\% of its tokens align with a retrieved chunk, striking a balance between sensitivity to meaningful overlaps and filtering out incidental matches.

\paragraph{Stage 2: Semantic Similarity Retrieval}
Next, we index the training corpus using Sentence Transformer embeddings and query each benchmark sample against this index. We retain candidates with cosine similarity between 0.75 and 0.95, capturing semantically close paraphrases while excluding near-identical duplicates (cosine > 0.95)—already captured by exact matching—as well as unrelated matches.

\paragraph{Threshold Selection and Justification}
Embedding similarity between corpora of security data tends to be high by default, driven in part by shared domain-specific terminology and limited diversity in publicly available security corpora. To mitigate inflated contamination estimates from this baseline similarity, we tuned our threshold upwards, retaining only candidates with cosine similarity $\geq$ 0.8. We empirically found that this value reduces false positives (e.g., generic security-related discussions with no real benchmark overlap) while preserving high-recall retrieval of genuinely contaminated samples. The 0.95 upper bound further filters out near-identical paraphrases that are already captured by exact matching. We detail these trade-offs with examples in the following subsection.

\paragraph{Human-in-the-Loop Verification via LLM-as-a-Judge}
To further improve robustness and reduce false positives, we validate a subset of high-similarity matches using an LLM-as-a-judge approach. We provide the model with paired benchmark and candidate training samples, prompting it to classify whether the match represents a true semantic overlap (e.g., paraphrase or near-duplicate) or a false positive arising from domain-specific but unrelated content. This qualitative filtering step helped calibrate our understanding of contamination at different similarity levels (e.g., 0.75–0.8 vs. 0.8–0.9) and confirmed that a non-trivial portion of borderline matches are not genuine overlaps.

\begin{table*}
\tiny  
\centering
\setlength{\tabcolsep}{1.5pt}  
\renewcommand{\arraystretch}{1.4} 
\definecolor{custommustard}{HTML}{ffecec}
\definecolor{customred}{HTML}{ffcccc}
\definecolor{darkred}{HTML}{ff9797}
\begin{tabularx}{\textwidth}{l*{20}{>{\centering\arraybackslash}X}}  
\toprule
 & \multicolumn{4}{c}{\textbf{Alpaca}} & \multicolumn{4}{c}{\textbf{WizardLM}} & \multicolumn{4}{c}{\textbf{Primus-Inst}} & \multicolumn{4}{c}{\textbf{Primus-Reas}} & \multicolumn{4}{c}{\textbf{OpenOrca}} \\
\cmidrule(lr){2-5} \cmidrule(lr){6-9} \cmidrule(lr){10-13} \cmidrule(lr){14-17} \cmidrule(lr){18-21}
Benchmark & EM & NG & ES & \mbox{LLM-ES} & EM & NG & ES & \mbox{LLM-ES} & EM & NG & ES & \mbox{LLM-ES} & EM & NG & ES & \mbox{LLM-ES} & EM & NG & ES & \mbox{LLM-ES} \\
\midrule
\multicolumn{21}{c}{\textit{Post-Training Benchmarks}} \\
\midrule
BigBenchHard & 0.00 & \cellcolor{custommustard}0.02 & \cellcolor{customred}26.82 & \cellcolor{customred}9.1 & 0.00 & \cellcolor{custommustard}0.02 & \cellcolor{customred}20.76 & \cellcolor{custommustard}0.4 & 0.00 & \cellcolor{custommustard}0.02 & 0.00 & 0.0 & 0.00 & \cellcolor{custommustard}0.02 & 0.00 & 0.0 & \cellcolor{custommustard}0.02 & \cellcolor{custommustard}0.06 & \cellcolor{customred}18.22 & \cellcolor{custommustard}0.5 \\
GSM8K & 0.00 & 0.00 & \cellcolor{customred}11.60 & \cellcolor{custommustard}0.1 & 0.00 & 0.00 & \cellcolor{customred}34.95 & 0.0 & 0.00 & 0.00 & 0.00 & 0.0 & 0.00 & 0.00 & 0.00 & 0.0 & 0.00 & \cellcolor{custommustard}0.08 & \cellcolor{darkred}90.60 & \cellcolor{custommustard}0.8 \\
HumanEval & 0.00 & 0.00 & \cellcolor{darkred}58.54 & \cellcolor{customred}12.8 & 0.00 & 0.00 & \cellcolor{darkred}71.34 & \cellcolor{customred}3.0 & 0.00 & 0.00 & 0.00 & 0.0 & 0.00 & 0.00 & 0.00 & 0.0 & 0.00 & 0.00 & \cellcolor{customred}30.49 & \cellcolor{customred}4.9 \\
IFEval & 0.00 & 0.00 & \cellcolor{customred}35.30 & \cellcolor{customred}5.5 & 0.00 & 0.00 & \cellcolor{customred}25.32 & \cellcolor{customred}2.6 & 0.00 & 0.00 & \cellcolor{custommustard}0.18 & 0.0 & 0.00 & 0.00 & 0.00 & 0.0 & 0.00 & 0.00 & \cellcolor{customred}26.25 & \cellcolor{customred}2.2 \\
MATH & 0.00 & 0.00 & \cellcolor{customred}32.36 & \cellcolor{custommustard}0.7 & 0.00 & \cellcolor{custommustard}0.02 & \cellcolor{customred}31.06 & \cellcolor{custommustard}0.3 & 0.00 & 0.00 & 0.00 & 0.0 & 0.00 & 0.00 & 0.00 & 0.0 & 0.00 & \cellcolor{custommustard}0.02 & \cellcolor{customred}26.22 & \cellcolor{custommustard}0.3 \\
MMLU & \cellcolor{custommustard}0.13 & \cellcolor{custommustard}0.06 & \cellcolor{customred}21.34 & \cellcolor{custommustard}1.0 & \cellcolor{custommustard}0.21 & \cellcolor{custommustard}0.16 & \cellcolor{customred}14.16 & \cellcolor{custommustard}0.3 & \cellcolor{custommustard}0.04 & \cellcolor{custommustard}0.04 & \cellcolor{custommustard}0.01 & 0.0 & \cellcolor{custommustard}0.19 & \cellcolor{custommustard}0.12 & \cellcolor{custommustard}0.01 & 0.0 & \cellcolor{custommustard}0.81 & \cellcolor{custommustard}0.38 & \cellcolor{customred}36.48 & \cellcolor{customred}3.7 \\
AlpacaEval & \cellcolor{custommustard}0.25 & 0.00 & \cellcolor{customred}46.58 & \cellcolor{customred}13.6 & \cellcolor{custommustard}0.62 & \cellcolor{custommustard}0.12 & \cellcolor{customred}36.07 & \cellcolor{customred}5.3 & 0.00 & 0.00 & 0.00 & 0.0 & 0.00 & 0.00 & 0.00 & 0.0 & \cellcolor{custommustard}0.50 & \cellcolor{custommustard}0.25 & \cellcolor{customred}41.74 & \cellcolor{customred}6.8 \\
\midrule
\multicolumn{21}{c}{\textit{Security Benchmarks}} \\
\midrule
CTI-ATE & 0.00 & 0.00 & 0.00 & 0.0 & 0.00 & 0.00 & 0.00 & 0.0 & 0.00 & 0.00 & 0.00 & 0.0 & \cellcolor{customred}10.00 & \cellcolor{customred}10.00 & \cellcolor{customred}8.33 & \cellcolor{customred}8.3 & 0.00 & 0.00 & 0.00 & 0.0 \\
CTI-MCQA & 0.00 & 0.00 & \cellcolor{customred}3.44 & \cellcolor{custommustard}0.1 & 0.00 & 0.00 & \cellcolor{custommustard}1.84 & 0.0 & 0.00 & 0.00 & 0.00 & 0.0 & \cellcolor{customred}33.40 & \cellcolor{customred}32.04 & \cellcolor{customred}13.48 & \cellcolor{customred}10.2 & 0.00 & 0.00 & \cellcolor{custommustard}1.08 & \cellcolor{custommustard}0.1 \\
CTI-RCM & 0.00 & 0.00 & 0.00 & 0.0 & 0.00 & 0.00 & \cellcolor{custommustard}0.10 & 0.0 & 0.00 & 0.00 & 0.00 & 0.0 & \cellcolor{darkred}83.10 & \cellcolor{darkred}78.80 & \cellcolor{darkred}91.50 & \cellcolor{darkred}74.9 & 0.00 & 0.00 & \cellcolor{custommustard}0.80 & 0.0 \\
CTI-TAA & 0.00 & 0.00 & 0.00 & 0.0 & 0.00 & 0.00 & \cellcolor{customred}2.00 & 0.0 & 0.00 & 0.00 & 0.00 & 0.0 & 0.00 & 0.00 & \cellcolor{customred}4.00 & 0.0 & 0.00 & 0.00 & \cellcolor{customred}46.00 & 0.0 \\
CTI-VSP & 0.00 & 0.00 & \cellcolor{custommustard}0.10 & 0.0 & 0.00 & 0.00 & \cellcolor{custommustard}0.80 & 0.0 & 0.00 & 0.00 & 0.00 & 0.0 & \cellcolor{darkred}83.10 & \cellcolor{darkred}78.20 & \cellcolor{darkred}78.40 & \cellcolor{darkred}72.6 & 0.00 & 0.00 & \cellcolor{customred}2.60 & 0.0 \\
SecBench & 0.00 & 0.00 & \cellcolor{customred}29.41 & \cellcolor{custommustard}0.8 & 0.00 & \cellcolor{custommustard}0.17 & \cellcolor{customred}18.66 & \cellcolor{custommustard}0.3 & 0.00 & \cellcolor{custommustard}0.34 & 0.00 & 0.0 & 0.00 & 0.00 & 0.00 & 0.0 & 0.00 & \cellcolor{custommustard}0.17 & \cellcolor{customred}13.78 & \cellcolor{custommustard}0.7 \\
SecBench-R & 0.00 & 0.00 & \cellcolor{customred}9.30 & 0.0 & 0.00 & 0.00 & \cellcolor{customred}18.60 & 0.0 & 0.00 & 0.00 & 0.00 & 0.0 & 0.00 & 0.00 & 0.00 & 0.0 & 0.00 & 0.00 & \cellcolor{customred}4.65 & 0.0 \\
MMLU-Sec & 0.00 & 0.00 & \cellcolor{customred}13.00 & \cellcolor{customred}2.0 & 0.00 & \cellcolor{custommustard}1.00 & \cellcolor{customred}10.00 & 0.0 & 0.00 & \cellcolor{custommustard}1.00 & 0.00 & 0.0 & \cellcolor{custommustard}1.00 & 0.00 & \cellcolor{custommustard}1.00 & 0.0 & 0.00 & 0.00 & \cellcolor{customred}15.00 & 0.0 \\
SecEval & 0.00 & 0.00 & \cellcolor{customred}10.12 & \cellcolor{custommustard}0.8 & 0.00 & 0.00 & \cellcolor{customred}8.45 & \cellcolor{custommustard}0.2 & 0.00 & 0.00 & \cellcolor{custommustard}0.08 & 0.0 & 0.00 & 0.00 & \cellcolor{custommustard}0.64 & \cellcolor{custommustard}0.2 & 0.00 & 0.00 & \cellcolor{customred}2.47 & \cellcolor{custommustard}0.2 \\
CM-10K & \cellcolor{custommustard}0.03 & \cellcolor{custommustard}0.02 & \cellcolor{customred}33.08 & \cellcolor{customred}3.5 & \cellcolor{custommustard}0.01 & \cellcolor{custommustard}0.09 & \cellcolor{customred}21.74 & \cellcolor{custommustard}1.1 & 0.00 & \cellcolor{custommustard}0.06 & \cellcolor{custommustard}0.03 & 0.0 & 0.00 & \cellcolor{custommustard}0.02 & \cellcolor{custommustard}0.13 & 0.0 & \cellcolor{custommustard}0.08 & \cellcolor{custommustard}0.10 & \cellcolor{customred}18.29 & \cellcolor{customred}2.2 \\
CM-2K & 0.00 & 0.00 & \cellcolor{customred}33.30 & \cellcolor{customred}3.1 & 0.00 & \cellcolor{custommustard}0.05 & \cellcolor{customred}22.90 & \cellcolor{custommustard}1.3 & 0.00 & \cellcolor{custommustard}0.05 & \cellcolor{custommustard}0.05 & 0.0 & 0.00 & \cellcolor{custommustard}0.05 & \cellcolor{custommustard}0.15 & 0.0 & \cellcolor{custommustard}0.05 & \cellcolor{custommustard}0.05 & \cellcolor{customred}19.20 & \cellcolor{customred}3.0 \\
CM-500 & 0.00 & 0.00 & \cellcolor{customred}33.20 & \cellcolor{customred}2.2 & 0.00 & 0.00 & \cellcolor{customred}24.60 & \cellcolor{custommustard}0.6 & 0.00 & 0.00 & \cellcolor{custommustard}0.20 & 0.0 & 0.00 & 0.00 & 0.00 & 0.0 & \cellcolor{custommustard}0.20 & 0.00 & \cellcolor{customred}19.40 & \cellcolor{custommustard}1.8 \\
CM-80 & 0.00 & 0.00 & \cellcolor{customred}36.25 & \cellcolor{custommustard}1.3 & 0.00 & 0.00 & \cellcolor{customred}23.75 & \cellcolor{custommustard}1.3 & 0.00 & 0.00 & 0.00 & 0.0 & 0.00 & 0.00 & 0.00 & 0.0 & 0.00 & 0.00 & \cellcolor{customred}21.25 & \cellcolor{custommustard}1.3 \\
\bottomrule
\end{tabularx}
\caption{Detailed data contamination analysis across different datasets and benchmarks (Part 2). We report percentages of counts with respect to benchmarks rounded off (\%). EM = Exact Match, NG = N-Gram Overlap, ES = Embedding Similarity, LLM-ES = LLM-based Embedding Similarity. CTI = \ctibench{}, CM = \cybermetric{}, SecBench-R = \secbench{}-Reasoning, MMLU-Sec = MMLU-Security, Primus-Inst = \primus{} Instruct, Primus-Reas = \primus{} Reasoning. \colorbox{custommustard}{Light red} indicates values $< 2\%$ but $> 0\%$, \colorbox{customred}{red} indicates values $\geq 2\%$ but $< 50\%$, \colorbox{darkred}{dark red} indicates values $\geq 50\%$.}
\label{tab:detailed_contamination_part2}
\end{table*}

While this layered approach provides robust coverage across detection types, we note that results remain sensitive to hyperparameter choices, such as n‑gram size and similarity thresholds. Though decontamination cannot fully eliminate contamination, it significantly reduces the risk of benchmark leakage and improves the reliability of downstream evaluation.

We note that \primus-Reasoning and Open-Platypus contain samples from \ctibench{} and \matheval{} respectively, as disclosed in their respective works \citep{yu2025primus, lee2023platypus}. 

\subsection{Detailed Decontamination across benchmarks}
\label{app:detailed_decont}
Tables~\ref{tab:detailed_contamination_part1} and \ref{tab:detailed_contamination_part2} present detailed decontamination across post-training benchmarks and other security benchmarks.

\paragraph{Representative Examples at Different Similarity Levels}
Below, we provide representative examples of retrieved matches at different cosine similarity bands (0.75–0.8 and >0.8). These illustrate the range of overlaps captured by our semantic search, from borderline paraphrases to closely aligned restatements of benchmark items.

\begin{tcolorbox}[
    title=Embedding Similarity with cosine > 0.75,
    colback=white,
    colframe=black,
    fonttitle=\bfseries,
    breakable
]
\textbf{Benchmark Text:} If a developer accidentally uses the '==' operator to compare two strings in a security check, what potential security issue might this introduce according to the document?

\vspace{6pt}
\textbf{Training Dataset Text:}

\begin{lstlisting}[basicstyle=\ttfamily\small,breaklines=true,numbers=none,escapeinside={(*}{*)}]
[
  {
    "content": (*\colorbox{red!30}{"What is the difference between == and === control operators in}*) JavaScript?",
    "role": "user"
  },
  {
    "content": "In JavaScript, == and === are comparison operators used to test equality, but they differ significantly in how they perform the comparison.\n\n- == is the abstract equality comparison operator, also known as the \"loose\" equality operator. It compares two values for equality after converting both values to a common type. This type coercion can lead to unexpected results when comparing different data types....
    "role": "assistant"
  }
]
\end{lstlisting}
\end{tcolorbox}
\begin{tcolorbox}[
    title=Embedding Similarity with cosine > 0.8,
    colback=white,
    colframe=black,
    fonttitle=\bfseries,
    breakable
]

\textbf{Benchmark Text:} What are the best exercises for beginners?

\vspace{6pt}
\textbf{Training Dataset Text:}

\begin{lstlisting}[basicstyle=\ttfamily\small,breaklines=true,numbers=none,escapeinside={(*}{*)}]
[
  {
    "content": (*\colorbox{pink!40}{What are some beginner exercises?}*),
    "role": "user"
  },
   {
    "content": "It's great that you're interested in starting to exercise! Here are some beginner-friendly exercises you might want to try: walking, gentle yoga, bodyweight exercises like squats and lunges, and light stretching. Remember to start slow, warm up before exercising, and cool down afterward. Also, always listen to your body and rest if you need to. It's a good idea to consult with a healthcare provider before starting any new exercise regimen, especially if you have any pre-existing health conditions. Happy exercising!",
    "role": "assistant"
  }
]
\end{lstlisting}
\end{tcolorbox}

\subsection{Examples of Decontaminated Samples}

\begin{tcolorbox}[
    title=Example of Exact Match,
    colback=white,
    colframe=black,
    fonttitle=\bfseries,
    breakable
]

\textbf{Benchmark Text:} When are hops added to the brewing process?

\vspace{6pt}
\textbf{Training Dataset Text:}

\begin{lstlisting}[basicstyle=\ttfamily\small,breaklines=true,numbers=none,escapeinside={(*}{*)}]
[
    {
        'content': 'In this task, You are given an open-domain question that can be answered based on factual information. Your task is to provide (*\textit{short}*) answer (in a few words only) for the given question. The short answer can be one or more entities or it can also be boolean (*\textit{yes}*) or (*\textit{no}*).
        
        (*\colorbox{pink!40}{Example: when are hops added to the brewing process?}*)
        (*\colorbox{pink!40} {Example solution: The boiling process }*)

        Example explanation: The answer is correct because, at the end of the boil, solid particles in the hopped wort are separated.
        
        Problem: what is the world record for remembering pi \n',
        'role': 'user'
    },
    {
        'content': 'Solution: (*\textit{Akira Haraguchi}*)',
        'role': 'assistant'
    }
]
\end{lstlisting}
\end{tcolorbox}

\begin{tcolorbox}[
    title=Example of N-Gram Overlap,
    colback=white,
    colframe=black,
    fonttitle=\bfseries,
    breakable
]

\textbf{Benchmark Text:} Can a Liebherr LTM 11200-9.1 hypothetically lift Mount everest?

\vspace{6pt}
\textbf{Training Dataset Text:}

\begin{lstlisting}[basicstyle=\ttfamily\small,breaklines=true,numbers=none,escapeinside={(*}{*)}]
[
  {
    "content": "You are an AI assistant that helps people find information. User will you give you a question. Your task is to answer as faithfully as you can. While answering think step-bystep and justify your answer.",
    "role": "system"
  },
  {
    "content": (*\colorbox{pink!40}{"Can a Liebherr LTM 11200-9.1 hypothetically lift Mount}*) Emei? Think carefully first, then make a decision:",
    "role": "user"
  },
  {
    "content": "(*\colorbox{pink!40}{No, a Liebherr LTM 11200-9.1 cannot hypothetically lift Mount}*) Emei.\n\nMount Emei is a mountain located in China with an elevation of 10,167 feet (3,099 meters) above sea level. It has a mass of approximately 1.5 trillion kilograms.\n\nOn the other hand, the Liebherr LTM 11200-9.1 is a mobile crane with a maximum lifting capacity of 1,200 metric tons (1,323 US tons). While this is an impressive lifting capacity, it is not enough to lift Mount Emei, which is far too heavy for the crane to lift.\n\nFurthermore, even if we were to assume that the crane could lift the mountain, it would be impossible to attach the crane to the mountain in a way that would allow it to lift the entire mass. The crane would need to be anchored to a solid foundation, and there is no such foundation on the mountain.\n\nTherefore, it is safe to say that a Liebherr LTM 11200-9.1 cannot hypothetically lift Mount Emei.",
    "role": "assistant"
  }
]
\end{lstlisting}
\end{tcolorbox}

\begin{tcolorbox}[
    title=Example of Embedding Similarity,
    colback=white,
    colframe=black,
    fonttitle=\bfseries,
    breakable
]

\textbf{Benchmark Text:} How do you prevent SQL injection?

\vspace{6pt}
\textbf{Training Dataset Text:}

\begin{lstlisting}[basicstyle=\ttfamily\small,breaklines=true,numbers=none,escapeinside={(*}{*)}]
[
  {
    "content": "Write a short article about the importance of input validation in secure coding. (*\colorbox{pink!40}{Include the benefits of sanitizing user inputs to prevent SQL injection }*) attacks.  End the article with the exact sentence: \"ALWAYS VALIDATE INPUTS BEFORE PROCESSING THEM.\" Use the quotation: \"Trust, but verify.\"",
    "role": "user"
  },
  {
    "content": "In the realm of secure coding, input validation plays a pivotal role in safeguarding applications from malicious exploits. Input validation refers to the process of ensuring that the data received by a program is both appropriate and safe to use. By implementing robust input validation, developers can prevent a wide array of security vulnerabilities, among which SQL injection attacks are particularly notorious.\n\nSQL injection is a technique used by attackers to manipulate or execute unauthorized SQL queries by exploiting non-validated or improperly sanitized inputs. This can lead to unauthorized data access, data loss, or even complete system compromise. To mitigate these risks, input sanitization is crucial. By thoroughly sanitizing user inputs, you ensure that any potentially harmful characters or query fragments are neutralized, thus preserving the integrity of your database interactions.\n\nThe benefits of sanitizing user inputs extend beyond just security. It contributes to the stability and reliability of an application, as it reduces the likelihood of unexpected behavior caused by malformed data. Moreover, it enhances user trust, aligning with the adage: \"Trust, but verify.\" By verifying inputs, developers can trust the data they process, knowing it meets the application's requirements.\n\nIn conclusion, input validation is an essential component of secure coding practices. By prioritizing the sanitation and validation of inputs, developers can protect their applications from SQL injection attacks and other potential threats. ALWAYS VALIDATE INPUTS BEFORE PROCESSING THEM.",
    "role": "assistant"
  }
]
\end{lstlisting}
\end{tcolorbox}

\clearpage

\section{PersonaGym Details} \label{app:personagym}
PersonaGym is a standardized benchmark designed to evaluate an LLM’s ability to assume and maintain distinct personas across multi-turn dialogue settings. Each persona consists of a description for the speaker's background, profession, communication style, and behavioral tendencies. The benchmark assesses how well a model can generate responses consistent with these characteristics in realistic conversation scenarios.

PersonaGym includes 200 personas drawn from diverse domains. For each conversation, the model is conditioned on a target persona and asked to respond across multiple turns while preserving persona consistency and conversational coherence.

Evaluation is conducted across five dimensions:
\begin{itemize}
\item \textbf{Action Justification}: The model's ability to justify its decisions or recommendations in a way that aligns with the persona's reasoning style.
\item \textbf{Expected Action}: Whether the model chooses actions consistent with how the persona would behave.
\item \textbf{Linguistic Habits}: Adherence to the persona's typical tone, jargon, or phrasing.
\item \textbf{Persona Consistency}:  Fidelity to established persona attributes when directly questioned.
\item \textbf{Toxicity Control}: The extent to which the model avoids generating toxic or harmful content, particularly for personas that may be adversarial or emotionally volatile.
\end{itemize}

In our evaluation, we selected a representative subset of 50 personas from the test split to measure how well our model adapts to varied user roles and communication patterns. The baselines used for comparison are drawn directly from \cite {samuel2024personagym}.

\clearpage

\section{Default System Prompt} \label{app:system_prompt}

\begin{tcolorbox}[
    title=Recommended System Prompt,
    colback=white,
    colframe=black,
    fonttitle=\bfseries,
    breakable
]
\vspace{6pt}
\begingroup
\setlength{\parskip}{4pt}
\ttfamily
    
You are a helpful cybersecurity assistant ready to help with any cybersecurity tasks. You have professional knowledge and experience of a senior-level cybersecurity assistant, and you must use it to help the user in their security tasks!

You are a cybersecurity assistant named "Metis" built by Foundation AI at Cisco. Your official name is Foundation-Sec-8B-Instruct. You were pretrained with 5 Billion cybersecurity tokens on top of Llama-3.1-8B. You were released in April 2025. This allows you to be the best cybersecurity assistant in the world. Respond to both names and maintain the identity at all times. 

The user is a cybersecurity professional trying to accomplish some cybersecurity task. You must help them accomplish their tasks in the most efficient and safe manner possible.

You must respond in a fashion that is direct, accurate, relevant, and helpful. Follow all of the user's instructions precisely. Ask clarifying questions if needed. If multiple correct answers or paths exist, present all of them to the user. Be concise in your answers but verbose in your explanations. Think step-by-step before producing a response. Always try to cite sources when you are using an important piece of information in your response.

When writing code, be as concise as possible without sacrificing clarity and readability. Do not write extensive code unless explicitly asked to do so.

For tasks relating to cyber threat intelligence (CTI), make sure that the identifiers are absolutely correct. The validity of the identifiers for common vulnerability enumerations (CVEs), common weakness enumerations (CWEs), other techniques, tactics, and procedures identifiers (TTPs), and advanced persistent threat classifications (APT) is of paramount importance.

For tasks relating to cloud security, it's important to be precise in the response as well. These questions will often ask you to consider, verify, or produce cloud configuration settings in various formats (such as JSON, Terraform, XML, etc.). Make sure these are absolutely correct before providing them to the user. Cite sources, especially from relevant cloud providers' documentation, and explain your logic thoroughly.

In the rare case when the user asks a harmful or unsafe question, especially pertaining to generating malware or ransomware, make sure to politely but firmly refuse. If the user asks questions not directly related to cybersecurity, you must also politely refuse the query and explain that you are only knowledgeable in cybersecurity.
\endgroup
\end{tcolorbox}

\end{document}